\documentclass[conference]{IEEEtran}
\usepackage{xcolor}
\IEEEoverridecommandlockouts
\usepackage{cite}
\usepackage{enumitem,amssymb}
\newlist{todolist}{itemize}{2}
\setlist[todolist]{label=$\square$}
\usepackage{dblfloatfix} 
\usepackage{amsmath,amssymb,amsfonts}
\usepackage{algorithmic}
\usepackage{graphicx}
\usepackage{textcomp}
\usepackage{listings}
\usepackage{float}
\usepackage{xcolor}
\usepackage{caption}
\usepackage{subcaption}
\usepackage{comment}
\usepackage{todonotes}
\usepackage{balance}
\usepackage{dblfloatfix}
\usepackage{makecell, caption, booktabs, multirow, siunitx}
\newcommand\tab[1][1cm]{\hspace*{#1}}
\usepackage[hyphens]{url}
\usepackage{hyperref}
\def\BibTeX{{\rm B\kern-.05em{\sc i\kern-.025em b}\kern-.08em
    T\kern-.1667em\lower.7ex\hbox{E}\kern-.125emX}}

\usepackage{nomencl}
\makenomenclature

\begin{document}
\title{Auto-SGCR: Automated Generation of Smart Grid Cyber Range Using IEC 61850 Standard Models}

\author{Muhammad M. Roomi, \IEEEmembership{Senior Member, IEEE}, S. M. Suhail Hussain, \IEEEmembership{Senior Member, IEEE}, \\ Ee-Chien Chang, David M. Nicol, \IEEEmembership{Fellow, IEEE} and Daisuke Mashima
}

\maketitle

\begin{abstract}
Digitalization of power grids have made them increasingly susceptible to cyber-attacks in the past decade. Iterative cybersecurity testing (i.e., red-team testing or penetration testing) is indispensable to counter emerging attack vectors and to ensure dependability of critical infrastructure. Furthermore, these can be used to evaluate cybersecurity configuration, effectiveness of the cybersecurity measures against various attack vectors, as well as to train smart grid cybersecurity experts defending the system. Enabling extensive experiments narrows the gap between academic research and production environment. A high-fidelity cyber range (a virtual cybersecurity testbed emulating smart grid systems) is vital as it is often infeasible to conduct such experiments and training using production environment. However, the design and implementation of cyber range requires extensive domain knowledge of physical and cyber aspect of the infrastructure. Furthermore, costs incurred for setup and maintenance of cyber range are significant. Moreover, most existing smart grid cyber ranges are designed as a one-off, proprietary system, and are limited in terms of configurability, accessibility, portability, and reproducibility. To address these challenges, an automated Smart grid Cyber Range generation framework (Auto-SGCR) is presented in this paper. Initially a human-/machine-friendly, XML-based modeling language called Smart Grid Modeling Language (SG-ML) was defined, which incorporates IEC 61850 System Configuration Language (SCL) files. Subsequently, a toolchain to parse SG-ML model files and automatically instantiate a functional smart grid cyber range was developed. The developed SG-ML models can be easily shared and/or modified to reproduce or customize for any cyber range. The application of Auto-SGCR is demonstrated through case studies with large-scale substation models. The toolchain along with example SG-ML models have been open-sourced. 

\end{abstract}


\pagestyle{plain}
\thispagestyle{plain}

\section{Introduction}
\label{sec:introduction}
The introduction of advanced information and communication technologies (ICT), integration of renewable energy sources, decentralisation of control, and deregulation of energy markets have led the conventional power systems to experience rapid transformation and evolve as smart grids. In this process, communication plays a pivotal role in smooth operation of the smart grid~\cite{ANCILLOTTI20131665}. As such, it is critical to define the standardized communication protocol to deliver reliable, secure, interoperable and cost-effective communication design for smart grids. Interoperability in communication design is an important feature that enables plug-and-play operation of different components in smart grid. In this regard, IEC 61850 standard has emerged as the most popular standard due to its object-oriented modeling approach for semantics and interoperable design~\cite{aftab20}. 

Standardization of communication, integration of IT technologies, and connectivity to external systems have widened the cyber-attack surfaces of smart grid systems, thus making the power grid an attractive target for cyber-attacks~\cite{SUN201845}. Due to its criticality, cybersecurity framework for smart grid has been laid by organisations such as National Institute of Standards and Technology (NIST)~\cite{8731}. To counter emerging cyber threats for dependable power grid operation and reliable electricity services, there is a need to continuously assess vulnerability, evaluate cyber-attacks impact, develop cybersecurity solutions and to train personnel for defending and restoring the system. For these purposes, an environment for conducting experiments and training is required. Given the real-time availability requirements and the risks of physical damages on power system infrastructure, it is not permissible to utilize real smart grid infrastructure for these evaluations. Alternatively, an isolated testbed using the same hardware as the real infrastructure can be used. However, such a testbed would incur high setup and maintenance cost and also reconfiguring or scaling-up is not possible. More importantly, cyberattack experiments that would damage equipment or cause safety hazards (e.g., Aurora Generator Test, which destroyed a diesel generator~\cite{idaho}) is not permissible.

In these regards, a cyber-physical range (or cyber range) is an effective alternative for studying cyber related issues of smart grids. Cyber range is a sandboxed, virtual environment that emulates physical and cyber configuration of smart grid systems. Some examples of cyber range in the literature is highlighted in~\cite{davis2013survey}. Cyber range allows researchers to test and experiment cyber-attacks with less constraints and to evaluate their impacts~\cite{yamin2020cyber}. Furthermore, new configurations could be tested in the cyber range to confirm the efficacy. As the smart grid cyber range should incorporate virtual power grid and virtual cyber system, design and configuration of these systems require intensive domain knowledge and engineering efforts. Numerous efforts are reported towards the development of cyber ranges. However, most of them are designed as one-off system for a specific physical/cyber system model and hence, not easily customizable. Conversely, many of them are proprietary system and hence, not publicly available or accessible. 

As a result, an open-source framework to flexibly configure and automatically instantiate IEC-61850-standard-based cyber range for smart grids, called {\it Auto-SGCR} is developed. This open-source framework enables a diverse range of users to customize the cyber range based on their requirements. Auto-SGCR consists of a modeling language, the Smart Grid Modeling Language (SG-ML) and a toolchain to process the SG-ML model files. SG-ML incorporates standardized configuration models, such as IEC 61850 SCL (System Configuration Language) files and IEC 61131 PLCopen XML files, which allows power grid operators to utilize their own configuration file to construct a cyber range. As SG-ML is XML-based, the toolchain can systematically parse and process it while human users can write, modify and interpret. This toolchain can be regarded as a `compiler' to translate the SG-ML model files into an operational cyber range. It is to be noted that the Auto-SGCR toolchain and the generated cyber range consists of open-source software. While commercially available software might have superior performance and fidelity, their adoption requires higher cost outlays and restrictions for public usage.The overview of the framework and preliminary results were published in~\cite{dsn2023}. 

This paper elaborates the definition and philosophy of SG-ML, and the processing logic and functionality of the toolchain to generate physical power system and cyber network topoplogy, and virtual smart grid devices (including, Intelligent Electronic Devices (IEDs), Programmable Logic Controllers (PLCs) and Supervisory Control and Data Acquisition Human Machine Interface (SCADA HMI). 

Case study using the proof-of-concept implementation of the toolchain with example SG-ML models are demonstrated in this paper. In summary, the major contributions of the paper are listed as follows:

\begin{itemize}
\item Elaborate the design and definition of SG-ML modelling framework that characterize the smart grid cyber range to be generated.
\item Detail the design and implementation of the Auto-SGCR toolchain for automated generation of smart grid cyber range. 
\item Case studies of cyber range generation and cyber-attack experiments for demonstrating practicability of Auto-SGCR.
 
\end{itemize}

Rest of the paper is organised as follows. Related works are explained in Section~\ref{related}. Section~\ref{overview} presents the configurations and model files required for the proposed Auto-SGCR framework to automatically generate the cyber range using IEC 61850 SCL files. Section~\ref{design} describes the  framework and the design of the proposed Auto-SGCR toolchain. The demonstration of the toolchain is discussed in Section~\ref{demonstration}, followed by cyber-attacks case studies in Section~\ref{casestudy}. Finally, Section~\ref{conclusion} presents the conclusions.

\begin{table*}[t]
\centering
\caption{Comparison of different co-simulation platforms for smart grid cyber security studies}
\begin{footnotesize}
  \begin{tabular}{|p{0.5cm}|p{1.2cm}|p{1.5cm}|p{2cm}|p{2.5cm}|p{2cm}|p{1.5cm}|p{1cm}|p{1.3cm}|}

    \hline
    \textbf{Ref}	& \textbf{Name} & \textbf{Power simulator} & \textbf{Network simulator/ emulator} & \textbf{Other Components} & \textbf{Protocols supported} & \textbf{Cyber-attack capabilities} & \textbf{License type} & \textbf{Automated configuration}\\
    \hline
    ~\cite{hardwarecps} & Resilient Energy Systems Lab (RESLab) & Power World Dynamic Studio (PWDS) & Common Open Research Emulator (CORE) & SEL RTAC & DNP3 & MITM, DoS & Proprietary & No\\
     \hline
    ~\cite{cosim3}& SCADA Cyber Security Testbed & PowerWorld & The Real-time Immersive Network Simulation Environment for Network Security Exercises (RINSE) & VPN clients & ModbusTCP & DDoS & Proprietary & No\\
    \hline
   ~\cite{2019817} & Offline co-simulation testbed & PSCAD & OMNeT++ & MATLAB for energy management system (EMS), SQlite & -- & DoS, FDI & Proprietary & No\\
    \hline
    ~\cite{cosim5} & -- & Opal RT & OPNET & -- & UDP & MITM & Proprietary & No\\
    \hline
    ~\cite{cosim6} & -- & DIgSILENT PowerFactory & OMNeT++ & Matlab for EMS & TCP/IP & Integrity and availability attacks & Proprietary & No\\
    \hline
    ~\cite{cosim7} & -- & RTDS & CORE & FreeOPCUA & OPC UA & DoS, MITM, Coordinated attack & Proprietary & No\\
    \hline
    ~\cite{cosim8} & -- & MATLAB & C++ in visual Studio and OPNET & -- & UDP & DoS, MITM, communication link outage & Proprietary & No\\
    \hline
    ~\cite{cosim9} & -- & GridLAB-D & NS-3 & HELICS, virtual machine emulator (QEMU), Opendnp3 & DNP3 & FDI and FCI attacks, command delay attacks & Proprietary & No\\
    \hline
    ~\cite{cosim10} & EPICTWIN & MATLAB/ Speedgoat & -- & IEDs, PLC, virtual switch as n headless Ubuntu 18.04.5 LTS virtual machine & IEC 61850 MMS, GOOSE & MITM & Proprietary & No\\ 
    \hline
    This work & Auto-SGCR & PandaPower & Mininet & Virtual IEDs, ScadaBR, OpenPLC61850 & IEC 61850 (MMS, GOOSE, SV, R-GOOSE, R-SV), ModbusTCP & MITM, FCI and FDI attacks & Fully Open Source & Yes\\ 
    \hline
  \end{tabular}
  \label{tabcosim}
\end{footnotesize}
\vspace{-3mm}
\end{table*}

\section{Related Works}\label{related}

In literature, many efforts for designing cyber-physical testbeds and digital twins for smart grids are reported~\cite{cpssurvey}. One such testbed is deployed in the University of Arkansas, which was designed mainly for research tasks on the detection of False Data Injection (FDI) attacks and vulnerability analysis of the Distributed Energy Resources (DER) cyber security schemes~\cite{DERtestbed}. The Institute for the Protection and Security of the Citizen in Italy also designed an experimental testbed for cyber vulnerability studies and countermeasures for enhancing security of the Supervisory Control And Data Acquisition (SCADA) system in the power system~\cite{scadatestbed}. The National SCADA testbed at the Idaho National Laboratory (INL) is also being used extensively for several research studies on vulnerability and impact analysis~\cite{scadaidaho}. The Electric Power and Intelligent Control (EPIC) testbed at Singapore University of Technology and Design (SUTD)~\cite{epic} has been developed to uncover potential integrity vulnerabilities in electrical synchronous generators. However, these physical testbeds are implemented with one specific system configuration and a fixed set of functionalities (e.g., for protection and automated control). This implies that the scalability, feasibility, and thereby, the scope of possible experiments conducted on the testbed is inherently limited. Furthermore, high-risk experiments, which could damage physical equipment, are prohibited. In addition, setting up physical testbeds is quite expensive owing to costs for hardware setup, and the accessibility to the physical testbeds is highly limited.

To address these issues, researchers have focused on developing smart grid cyber-physical range with a low-cost hardware and software for cyber security research and education~\cite{lowcost}. In~\cite{2017124}, a realistic cyber-physical environment where power system events are simulated in Real Time Digital Simulator (RTDS) is proposed. In~\cite{5759169}, a testbed has been developed using PowerWorld software~\cite{pw} for power system simulation and Modbus protocol to detect cyber-attacks on SCADA system. Similarly, in~\cite{fdimss}, cyber-attacks on Modbus protocol in SCADA system in a substation are studied where Pandapower was utilized to simulate the power system. Works in~\cite{hardwarecps},~\cite{2019817},~\cite{7116592},~\cite{cosim3},~\cite{cosim5},~\cite{cosim6},~\cite{cosim7},~\cite{cosim8},~\cite{cosim9} present testbeds or cyber ranges involving co-simulation platforms where power system tools are interfaced with network simulation tools. These works utilized combination of network simulation tools (such as OMNeT++, NS3, OPNET, etc.) and real device connectivity for simulating cyber network in smart grid cyber ranges. Table~\ref{tabcosim} shows the comparison of different smart grid cyber ranges involving co-simulation platforms. Most of the existing works utilize proprietary tools for power system and cyber network simulations making the platform hard to reproduce and reuse. Furthermore, these co-simulation platforms require additional procedures for setting up the configurations to simulate different scenarios.

Most of the cyber-physical ranges and testbeds reported in literature focused on SCADA ModbusTCP, DNP3, IEEE C37.118.2 synchrophasor protocols. However, limited works are reported on developing IEC 61850 communication based cyber ranges for smart grids even though IEC 61850 is a widely-implemented international standard. In~\cite{costCPS}, a software testbed for IEC 61850-based electrical substations based on open source software is proposed. Similarly, in~\cite{SoftGrid}, a software-based substation testbed `SoftGrid' for evaluating cyber security solutions is reported. However, these testbeds~\cite{costCPS},~\cite{SoftGrid} are limited to substation domain and does not extend to entire smart grid domain.  

In order to address these challenges towards configurable, extensible, and portable smart grid cyber range, Auto-SGCR enables automated generation of smart grid cyber range based on user-defined models. 
SG-ML (a human-/machine-friendly, cyber-range-modeling language based on standard IEC 61850 SCL) is designed and the tool for instantiation and deployment of cyber range is developed. SG-ML models can be easily shared and customized without intensive programming expertise and can be compiled into functional cyber ranges on user's environment with minimal engineering efforts.

\section{SG-ML: Smart Grid Modeling Language Framework}\label{overview}

In this section, the configurations of a cyber range that are necessary to emulate the physical-cyber behaviours of smart grid systems is discussed. Subsequently, the framework for modeling and instantiating such a smart grid cyber range by utilizing IEC 61850 SCL files is elaborated.

\subsection{Configurations for smart grid cyber range}\label{sec:sgcr-overview}

\begin{figure}[t]
\centerline
{\includegraphics[width=1\linewidth, height=3.5cm]{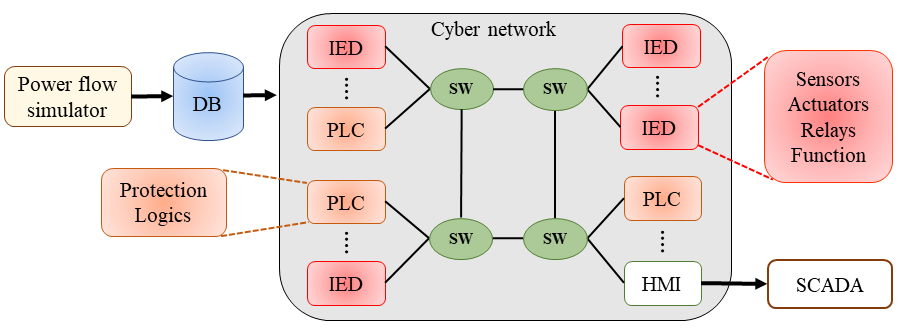}}
\caption{Smart Grid Cyber Range Architecture.}
\label{arch}
\vspace{-3mm}
\end{figure}

Fig.~\ref{arch} depicts the architecture of the smart grid cyber range and the virtual components to be deployed on the cyber range. At the high level, smart grid cyber range consists of: i) power flow simulator for calculating power system's physical parameters under specific system condition, and ii) cyber system emulator that consists of virtualized network equipment (e.g., switching hubs), virtualized smart grid devices, e.g., IEDs, PLCs, and SCADA HMI. IEDs interact with the power system simulator through database (e.g., to obtain measurements and circuit breaker (CB) status), and measurements are sent to the SCADA HMI and/or PLCs. PLCs may mediate communication between IEDs and SCADA HMI and execute automated control logic based on the measurements received from IEDs. Automated protection logic (e.g., over voltage/current protection) can also be implemented on IEDs. The interaction between the power simulator and the cyber emulator in near real-time is achieved through the database. Using such a smart grid cyber range, state-of-the-art cyber-attacks like False Command Injection (FCI) and FDI, as demonstrated in~\cite{fdimss}, ~\cite{roomi2023analysis} can be experimented. 

In summary, the design of such a smart grid cyber range requires the following configurations.   

\begin{itemize}
    \item Power system topology and configuration, e.g., Single Line Diagram (SLD), configuration and status of power system devices like CBs, load profile, etc.
    \item Cyber network topology and configuration, e.g., connectivity among devices, network bandwidth, etc.
    \item Device configurations, e.g., network addresses, communication models, and/or functionalities of SCADA HMI, PLCs, and IEDs.
\end{itemize}

\subsection{SG-ML model files}

In order to develop automated generation framework for smart grid cyber range, the key step is to define a modeling framework for the configurations listed in Section~\ref{sec:sgcr-overview}, which should be human-/machine- friendly. Furthermore, a standardized schema is utilized for defining the configuration files. Consequently, these files can be repurposed and/or modified using the existing tools or software. As a result, the XML-based IEC 61850 SCL files are explored.

Based on the analysis, most of the physical and network configurations can be derived from IEC 61850 SCL files. SCL is the descriptive language defined by IEC 61850 for configuring electrical power utility systems. The main parts of an SCL file include `Header', `Process', `Substation', `Communication', `IED', `Datatype templates'. There are different types of SCL files, viz., System Specification Description (SSD), IED Capability Description (ICD), System Exchange Description (SED) and System Configuration Description (SCD), which serve different purposes. These IEC 61850 SCL files are the main components in building the physical and cyber topology of smart grids. The files are standardized XML files, and thus it can be easily parsed by software and can also by easily interpreted/edited by users. Therefore, these IEC 61850 SCL files are utilized to define modeling framework. This design decision allows users (e.g., power grid operators) to optimize the existing SCL files to generate cyber range corresponding to their operational systems. 

\subsubsection{Power System Components} \label{psc}
SSD files are used to define the physical topology of power system, e.g. defining the SLD. Each major part/component of smart grid such as a substation, DER plant, etc., (termed as `\textit{Process}' in IEC 61850 SCL terminology~\cite{61850-6}) is defined by a single SSD file. An SED file is used to define the electrical connectivity between multiple `\textit{Process}' (i.e. substations or DER plants). 

\subsubsection{Cyber Network Components}\label{cnc}
Similar to power system, cyber network components are defined through ICD and SCD files. ICD files define device-specific information such as network addresses, supported protocols, etc and SCD files include details for defining process' (i.e substations'/DERs') local area network topology. These information are included in the `Communication' section in the SCL file. SED files are also used to define communication model, if any, between processes.

\subsubsection{Smart Grid Devices}
\label{vc}
One of the vital smart grid devices that is to be involved in cyber range is an IED. ICD files are utilized to instantiate virtual IEDs. In addition to the `Communication' section, ICD files include `IED' section that defines datasets, and implemented functionalities (e.g., type of protection functions). 

The aforementioned SCL files are the core components in configuring an operational cyber range. However, IEC 61850 SCL solely is not sufficient to generate functional cyber range. One of the limitations is that SCL (SSD) defines only the static configurations of power systems and lack information that is necessary to define dynamic configurations (for instance load profile - to run the power system simulation). Furthermore, SCL (ICD) files do not define the details of protection logic (e.g., threshold values) or exact mapping between IEC 61850 attribute names to physical power grid components and/or measurement. Automated control logic for PLCs, and data point configurations for SCADA HMI are also not defined in the SCL. In these regards, additional configurations are required and the details are elaborated in the following subsection.

\subsection{Supplementary Configuration XML Files}
Although SSD files are sufficient to understand the physical layout of any system, many necessary parameters of the physical components, such as generators, transformers, loads, etc. are not defined in these files. The attributes that can be fetched from the SCL files and the missing attributes that are additionally required to successfully develop a cyber range using Auto-SGCR are tabulated in Table~\ref{tab1}. In the table, the missing attributes are marked with `-' in `IEC 61850 SCL attribute' column. Therefore, a supplementary configuration file is defined using XML schema (named as \textit{Supplementary XML}) to provide these missing details. This XML file includes the parameter details such as generator ratings, transformer ratings, load profile, CB status at each time step, and the cable data. The details of the Supplementary XML file are shown in Fig.~\ref{ssdprop}. As depicted in the figure, the XML file includes details such as the location of the component, real and reactive power ratings of the power sources and loads, the status (open/close) of the CBs, and the cable parameters like its type and length in kms. Furthermore, the `data sequence' attribute of each element can be used to define the time-series.

Similarly, ICDs that are used to define the capabilities of IEDs does not include the address mapping and the threshold parameters. Therefore, two Supplementary XML files: `CPMapping XML' and `Thresholds XML' are defined. CPMapping XML is used to define mapping between power flow simulator output and the IEC 61850 information models. Using entries in CPMapping XML, information exchanged with IEC 61850 based communication is associated with the physical measurement and device status in the simulated power grid model. Similarly, Threshold XML defines threshold parameters for the protection functions (e.g., trip threshold) that are not found in the ICD files. A detailed explanation is provided in Section~\ref{virtualied}.

Additionally, to incorporate the definition of automated control logic that run on PLCs in the modeling framework, standardized PLCopen XML schema, which is defined as part of IEC 61131-3 standard~\cite{plcopen}, is incorporated in SG-ML. As SCADA attributes such as data sources and data points are not included in SCL, these are defined in `SCADA Config XML'.

In summary, SG-ML model files consists of IEC 61850 SCL files (SSD, SCD, ICD, and SED files) and the supplementary XML files described in the previous sections.

\begin{table*}[t]
\caption{Attributes in SCL for Power Components.}
\begin{center}
\renewcommand{\arraystretch}{1.1}
\begin{footnotesize}
\begin{tabular}{|p{2.8cm}|p{3.7cm}|p{10.1cm}|}
\hline

\textbf{\makecell{Components in\\ powerflow simulator}} & \textbf{\makecell{Processor attribute\\ (description)}} & 
\textbf{\makecell{IEC 61850 SCL attribute}} \\
\hline

\multirow{4}{*}{Generator} & 
\textit{‘bus’} (location of the generator) & 
\begin{lstlisting}[aboveskip=-6pt, belowskip=-10pt]
<ConductingEquipment name="" Type="GEN"> 
 <Terminal connectivityNode=""/>
</ConductingEquipment> 
\end{lstlisting}
\\
\cline{2-3} 
 & 
\textit{‘vn\_kv’} (grid voltage level) & 
\begin{lstlisting}[aboveskip=-6pt, belowskip=-10pt]
<Substation>
 <VoltageLevel name= numPhases="" nomFreq=""/>
</Substation>
\end{lstlisting}
\\
\cline{2-3} 
 & 
\textit{‘p\_mw’}(rated active power) & 
-\\
\cline{2-3} 
 & 
\textit{‘vm\_pu’}(rated voltage) & 
-\\
\cline{1-3} 

\multirow{3}{*}{PV/Battery} & 
\textit{‘bus’} (location of PV/Battery) & 
\begin{lstlisting}[aboveskip=-6pt, belowskip=-10pt]
<ConductingEquipment name="" Type="GEN">   
 <Terminal connectivityNode=""/>
</ConductingEquipment>
\end{lstlisting}
\\
\cline{2-3} 
& 
\textit{‘p\_mw’} (rated active power) & 
-\\
\cline{2-3} 
& 
\textit{‘q\_mvar’} (rated reactive power) & 
-\\
\cline{1-3} 

\multirow{2}{*}{Transformer} & 
\textit{‘hv\_bus’} (location of primary side of the transformer) & 
\begin{lstlisting}[aboveskip=-6pt, belowskip=-10pt] 
<PowerTransformer name="" type="PTR">
 <TransformerWinding name="" type="PTW">
  <Terminal connectivityNode="" voltageLevelName=""/>
 </TransformerWinding>
</PowerTransformer>
\end{lstlisting}
\\
\cline{2-3}
& 
\textit{‘lv\_bus’} (location of secondary side of the transformer) & 
\begin{lstlisting}[aboveskip=-6pt, belowskip=-10pt] 
<PowerTransformer name="" type="PTR">
 <TransformerWinding name="" type="PTW">
  <Terminal connectivityNode="" voltageLevelName=""/>
 </TransformerWinding>
</PowerTransformer>
\end{lstlisting}
\\
\cline{1-3} 

\multirow{3}{*}{Load} & 
\textit{‘bus’} (location of the load) & 
\begin{lstlisting}[aboveskip=-6pt, belowskip=-10pt] 
<ConductingEquipment name="" Type="IFL">
 <Terminal connectivityNode=""/>
</ConductingEquipment>
\end{lstlisting}
\\
\cline{2-3} 
& 
\textit{‘p\_mw’} (rated active power) & 
-\\
\cline{2-3} 
& 
\textit{‘q\_mvar’} (rated reactive power) & 
-\\
\cline{1-3}  

\multirow{3}{*}{Circuit Breaker} & 
\textit{‘bus’} (location of the switch) & 
\begin{lstlisting}[aboveskip=-6pt, belowskip=-10pt]
<ConductingEquipment name="" Type="CBR">
 <Terminal connectivityNode=""/> 
</ConductingEquipment>
\end{lstlisting}
\\
\cline{2-3} 
& 
\textit{‘closed’} (position of the switch) & 
-\\
\cline{2-3} 
& 
\textit{‘element’} (type of switch) & 
\begin{lstlisting}[aboveskip=-6pt, belowskip=-10pt] 
<ConductingEquipment name="" Type="CBR/DIS">
</ConductingEquipment>
\end{lstlisting}
\\
\cline{1-3}

\multirow{4}{*}{Cable} & 
\textit{‘from\_bus’} (cable starting node) & 
\begin{lstlisting}[aboveskip=-6pt, belowskip=-10pt] 
<Terminal connectivityNode=""/>
\end{lstlisting}
\\
\cline{2-3} 
& 
\textit{‘to\_bus’} (cable ending node) & 
\begin{lstlisting}[aboveskip=-6pt, belowskip=-10pt] 
<Terminal connectivityNode=""/>
\end{lstlisting}
\\
\cline{2-3} 
& 
\textit{‘length\_km’} (cable length) & 
-\\
\cline{2-3} 
& 
\textit{‘std\_type’} (cable type) & 
-\\
\cline{1-3}

\end{tabular}
\end{footnotesize}
\label{tab1}
\end{center}
\vspace{-3mm}
\end{table*}

\begin{figure}[t]
\centerline
{\includegraphics[width=\linewidth]{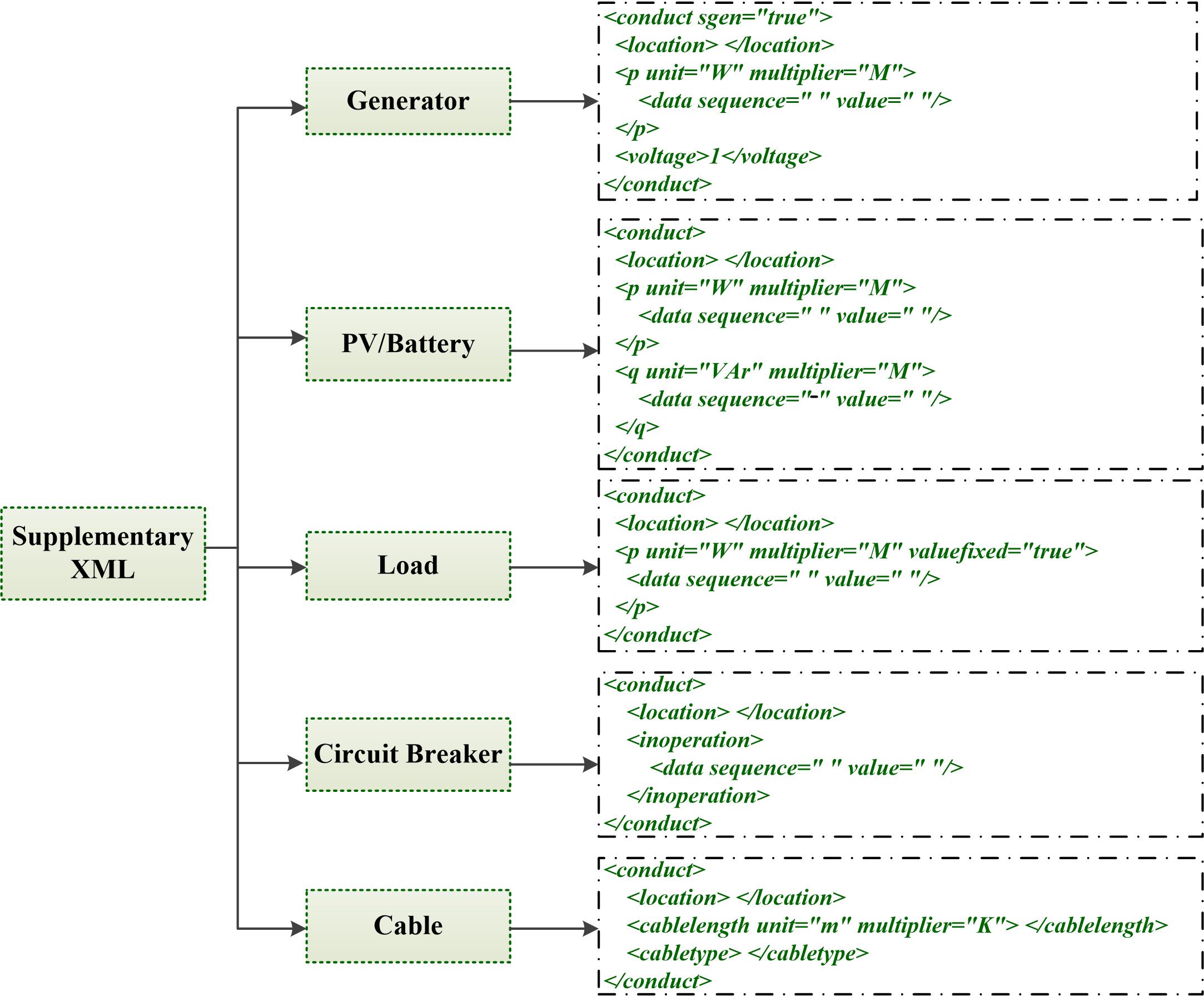}}
\caption{Supplementary XML for Power System Configuration.}
\label{ssdprop}
\end{figure}

\begin{figure}[!t]
\centerline
{\includegraphics[width=\linewidth]{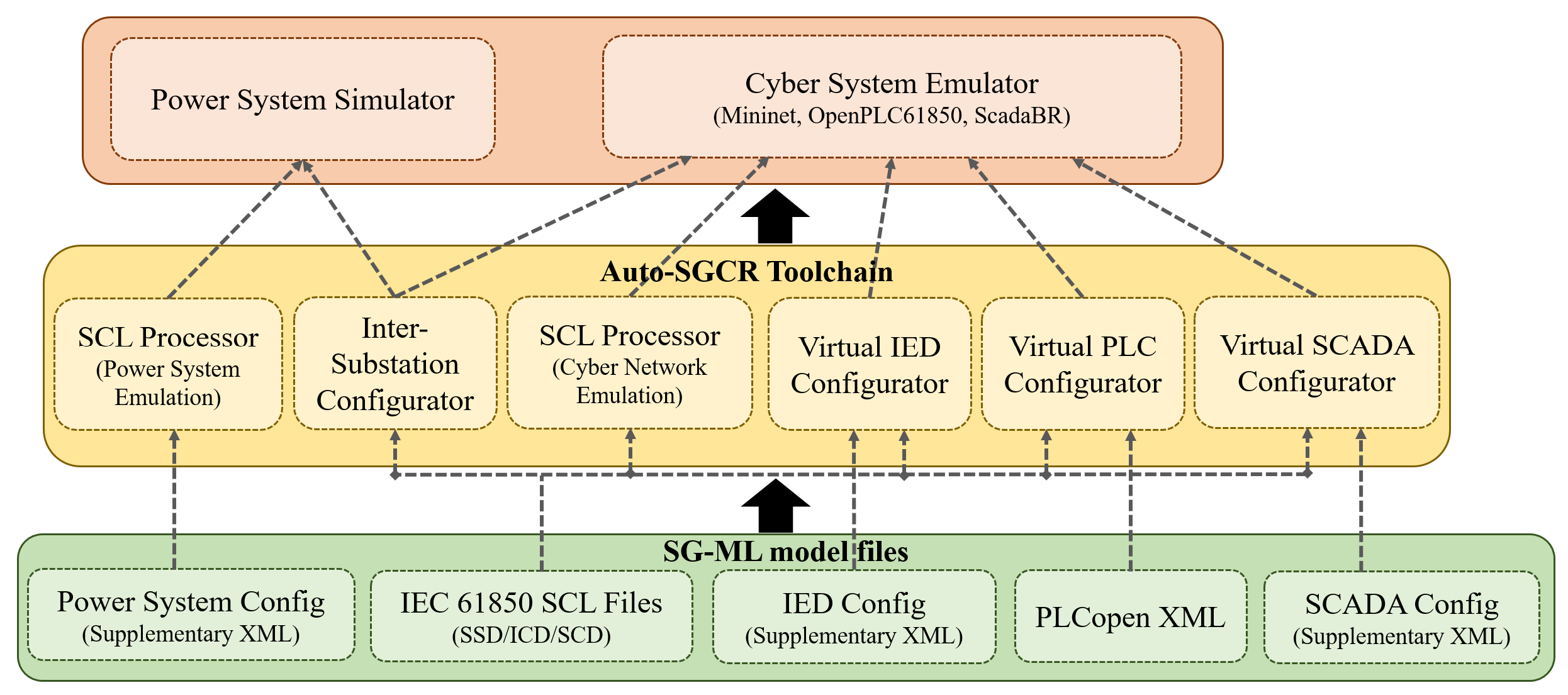}}
\caption{Auto-SGCR Framework.}
\label{sgmlframe}
\vspace{-3mm}
\end{figure}

\begin{figure}[t]
\centerline
{\includegraphics[width=0.8\linewidth, height=6cm]{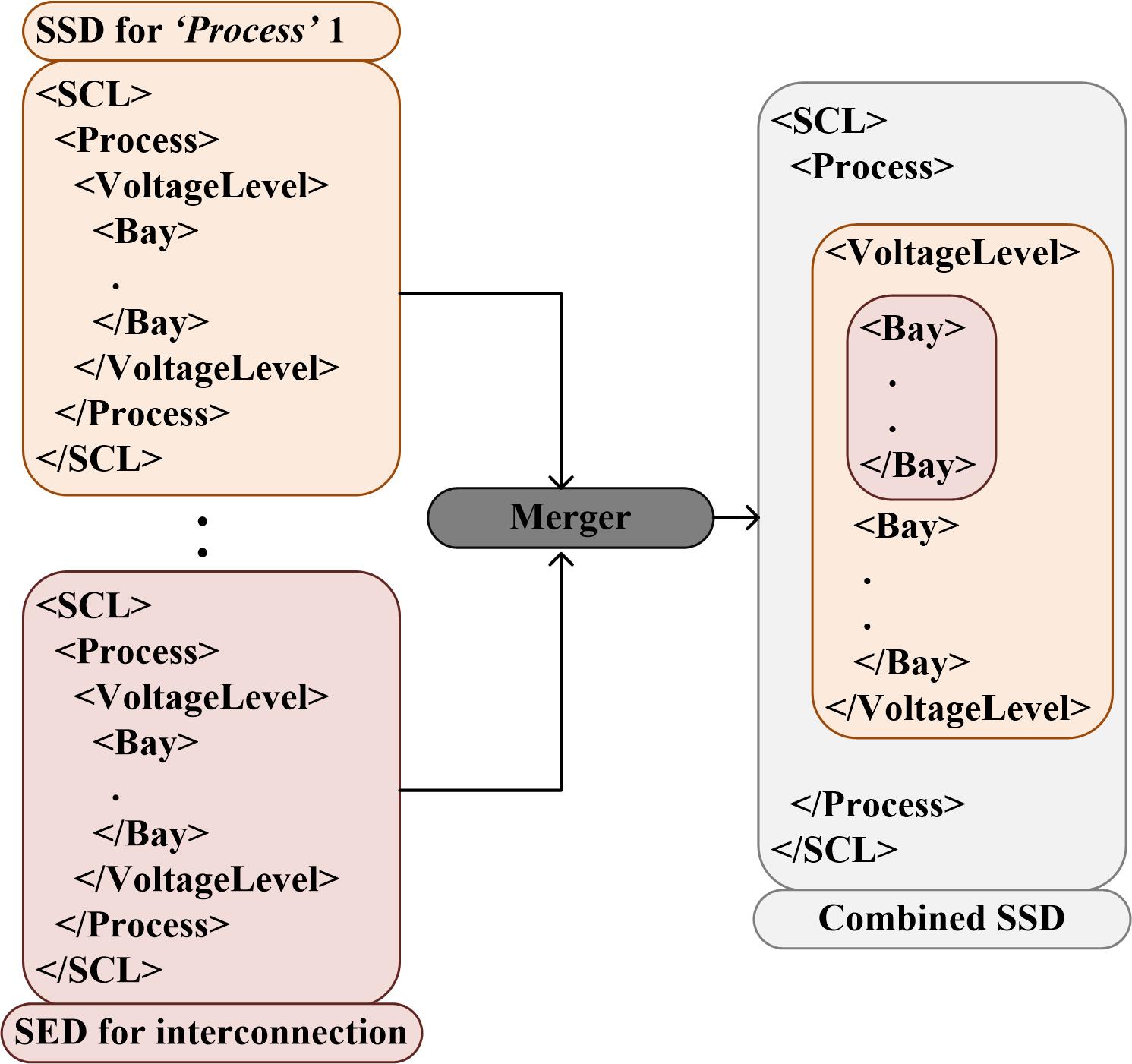}}
\caption{Combining Power System Components.}
\label{ssdsed}
\vspace{-5mm}
\end{figure}

\begin{figure}[t]
\centerline
{\includegraphics[width=0.8\linewidth, height = 6cm]{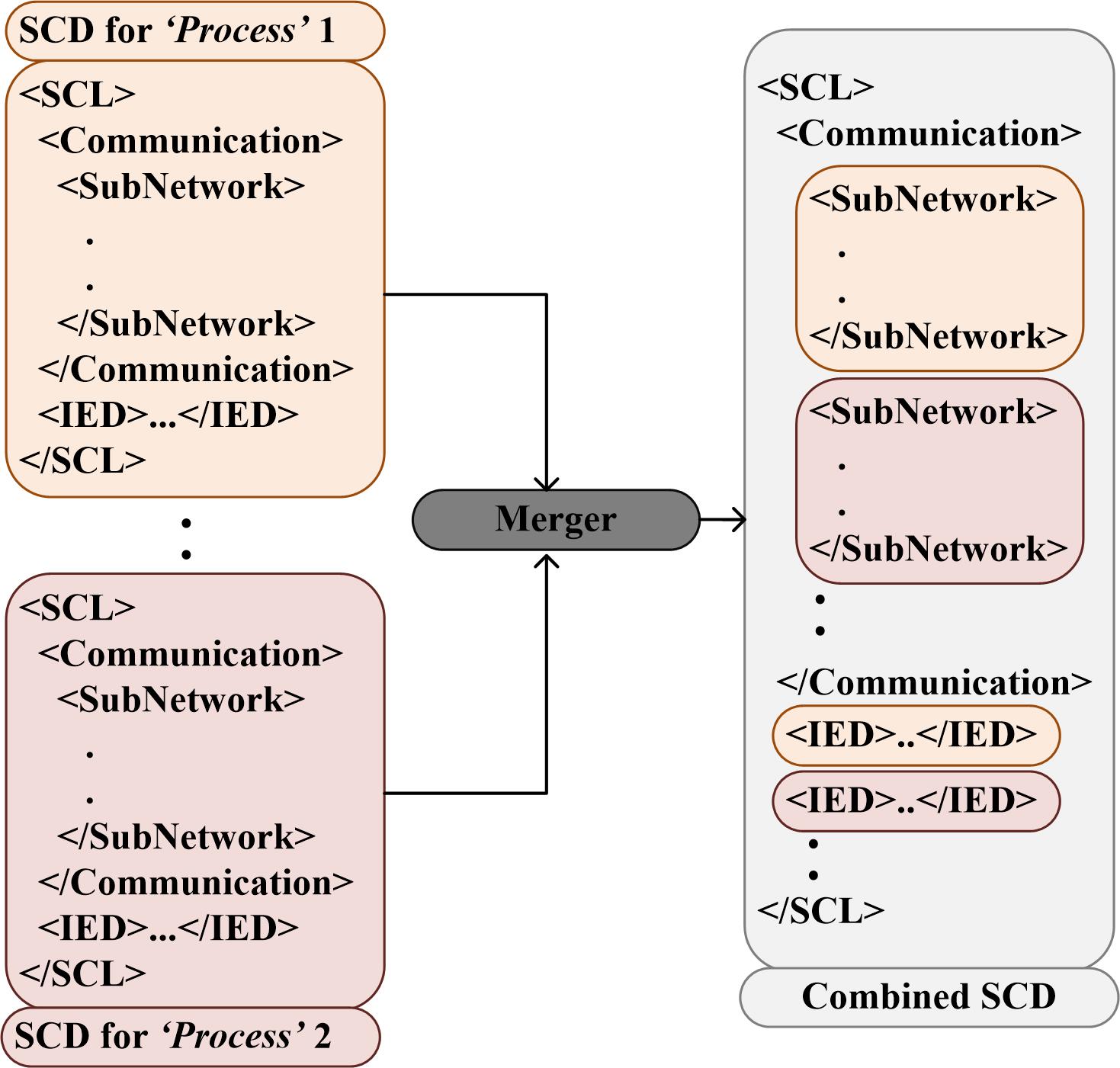}}
\caption{Combining Cyber Network Components.}
\label{scdscd}
\vspace{-5mm}
\end{figure}

\begin{figure*}[t]
\centerline
{\includegraphics[width=0.9\linewidth]{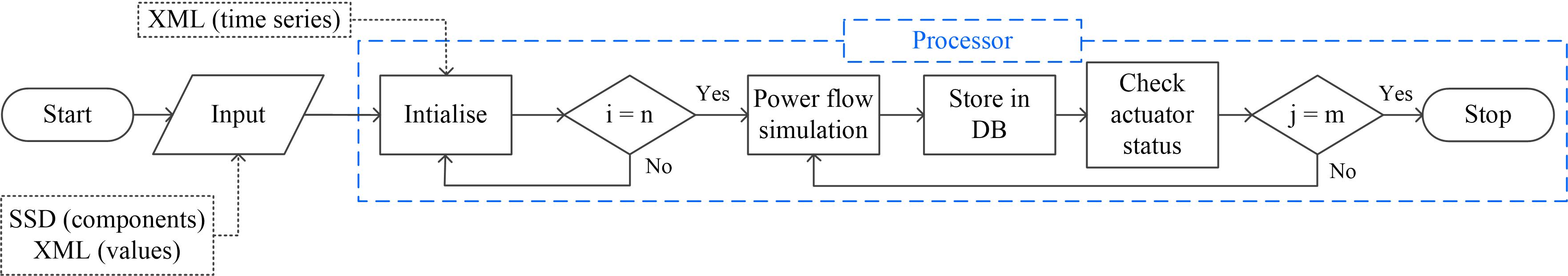}}
\caption{Flowchart of Configuration and Execution of Power Flow Simulation.}
\label{fcpsp}
\vspace{-5mm}
\end{figure*}

\section{Auto-SGCR Framework and Toolchain}\label{design}
The high-level architecture of the automated generation framework, Auto-SGCR, is illustrated in Fig.~\ref{sgmlframe}. The input SG-ML files are highlighted in green. The open-source tools for Auto-SGCR (highlighted in yellow) are implemented to process these modeling files to instantiate a cyber range, and details of the tool are discussed in this section. 

\subsubsection{Merger Tool}\label{merger}
In order to develop multi-substation cyber range using Auto-SGCR, an integration between the SSDs and the SEDs is necessary. As such, a merger tool is developed to merge the multiple SSD and SED files into a single merged SSD file. Fig.~\ref{ssdsed} portrays the SSD and SED sections in an SCL file that is utilized for creating a single SSD for cyber range. The SSD file typically consists of the `Process' section, `VoltageLevel' and `Bay' sub-sections. The merger concatenates `Process' section data from different SSD files to one final SSD file. The `VoltageLevel' tag in SED file is identified and its contents (i.e. `Bay' sub-sections) are merged under the similar `VoltageLevel' tag in final SSD file. This process is repeated for all the SED files. Thus, the final SSD file containing the complete electrical topology of entire grid is created. 

Similarly, the merger combines the different SCD files of each process (i.e., substations/DERs) to create a merged SCD file for large scale complete system. The merged SCD file is concatenation of different `SubNetwork' sections. Each `SubNetwork' section contains `Communication' section, which is extracted from the SCD file of each process. The information of all switches (i.e., `IED' elements) corresponding to different processes are listed in the merged SCD file. Fig.~\ref{scdscd} depicts the framework for combining different SCD files to one merged SCD file for entire grid. This merged SCD file is used to emulate the cyber-network topology of grids.

\subsubsection{Configuring Power System Simulator}\label{powersystememulation}

Auto-SGCR utilizes the `Process' section in any SSD files for emulating the power system topology. The ‘Process’ section in the SSD file includes different ‘Voltage Level(s)’, which in turn consists of ‘Bay(s)’. A single process may contain one or more voltage levels and a single voltage level may contain one or more bays. Voltage level includes details such as operating voltage, number of phases, operating frequency, alternating transformer voltages, etc. A ‘Bay’ represents a single feeder line in the process. Therefore, it includes the details of the physical components (represented as ‘ConductingEquipment’ in SCL) such as CBs, relays, disconnectors, and so forth, which are present in the feeder line. A node is the point of connection between two physical components, and is represented as ‘Terminal connectivityNode’ in the SCL file.

\begin{figure}[b]
\vspace{-3mm}
\centerline
{\includegraphics[width=0.95\linewidth, height=4cm]{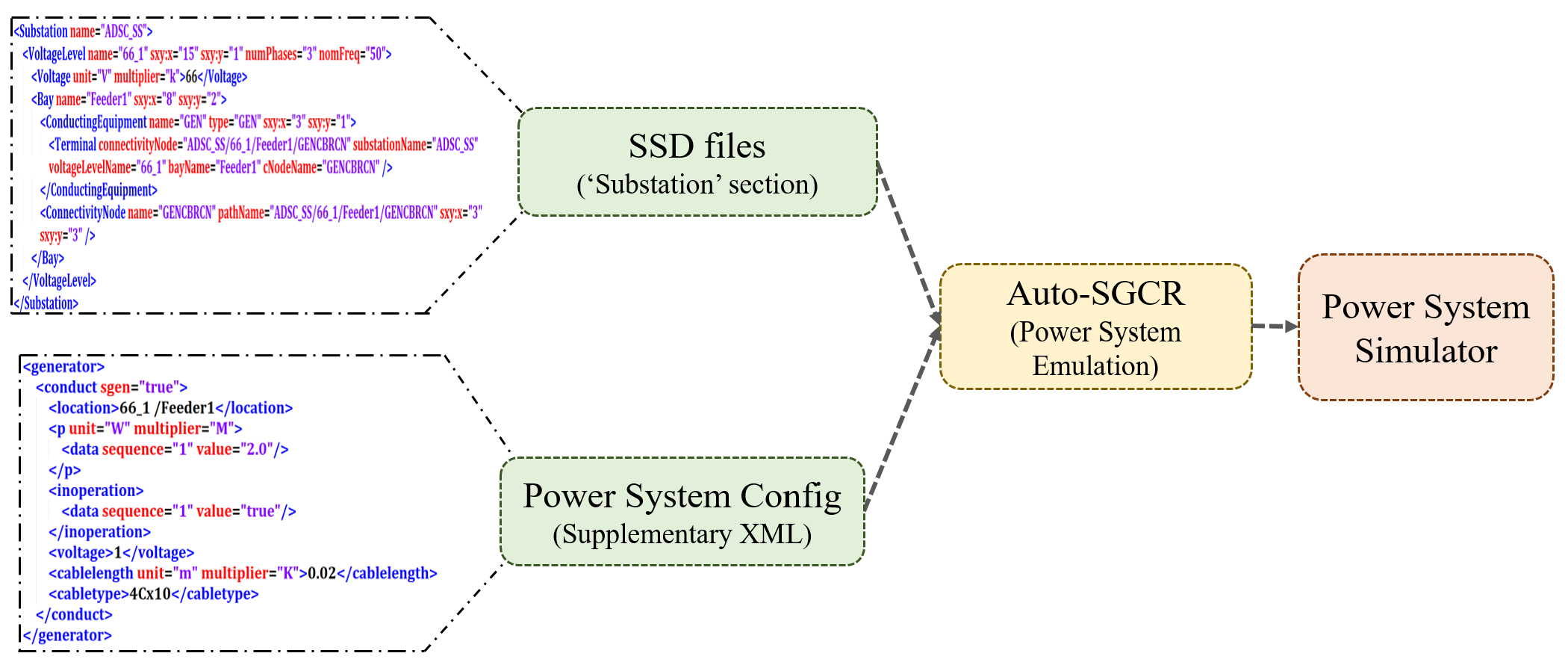}}
\caption{Physical System Implementation Framework.}
\label{ssdsgmlvsd}
\end{figure}

Auto-SGCR utilizes SSD and the Supplementary XML files to derive power system topology and then run power flow simulation. The flowchart of the power system simulation is depicted in (Fig.~\ref{fcpsp}). With the initialization of the whole power network components, the process of power flow simulation commences. The components are predefined in the SSD file and the file forms one part of the input. The rated values of the components are read from the Supplementary XML file. The initialization process generates a series of power flow simulation models according to the time-series definition in the Supplementary XML file. This time-series is defined using the `data sequence' attribute in the Supplementary XML. In this case, `$n$' time steps are defined in the Supplementary XML and thus, `$n$' power flow simulation models are generated. For the first time step, the power flow is calculated with the initial condition including the CB status stored in the database. Subsequently, the voltage, current and power measurements are stored in the database for the first iteration and these values are updated in the database for further iterations. Prior to an iteration, any changes in the CB status, which may be due to protection function execution or remote control by PLCs, SCADA HMI, or perhaps attackers, in the DB is checked and updated. Thereafter, the power system simulator in the smart grid cyber range utilizes the updated values for power flow calculation in the next iteration. This process is executed until all of the time steps are consumed. The overall framework for emulating the physical power system behaviour is depicted in Fig.~\ref{ssdsgmlvsd}. 


\begin{figure}
\centerline
{\includegraphics[width=0.95\linewidth, height=3.5cm]{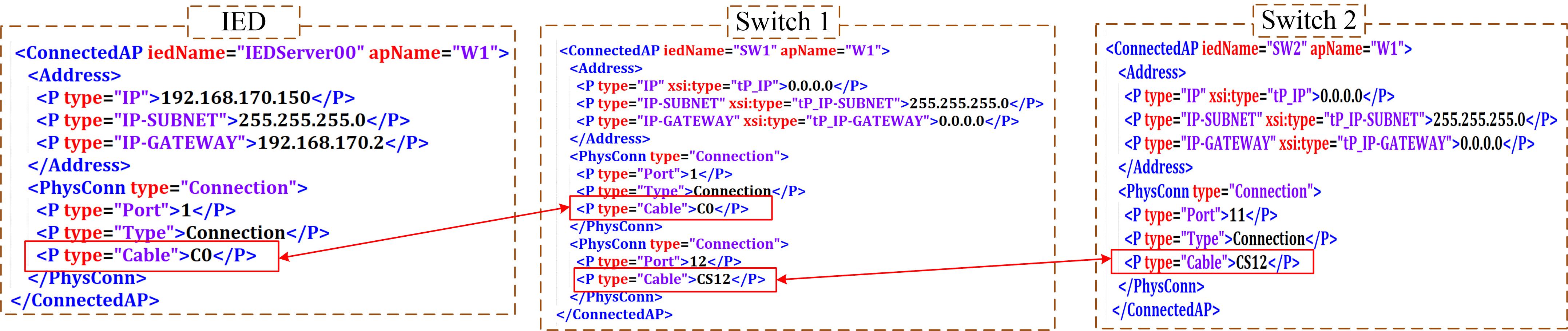}}
\caption{Cyber Network Connectivity Derivation for Network Emulator.}
\label{commemulation}
\vspace{-5mm}
\end{figure}

\begin{table*}[b]
\vspace{-3mm}
\centering
\renewcommand{\arraystretch}{1.1}
\caption{Protection Logics}
\begin{footnotesize}
  \begin{tabular}{|p{2cm}|p{2cm}|p{7cm}|p{3cm}|p{2cm}|}

    \hline
    \textbf{Classification}	& \textbf{Connectivity} & \textbf{Protections} & \textbf{Inputs} & \textbf{Protocols}\\
    \hline
    Single substation & Intra-substation &	Over-current (PTOC), Over-voltage (PTOV), Under-voltage (PTUV), Inter-tripping (PTRC), Inter-locking (CILO) & ICD, SCD, Supplementary XML & MMS, GOOSE\\

     \hline
   Multi-substations&	Intra-substation, Inter-substation&	PTOC, PTOV, PTUV, PTRC, CILO, Differential (PDIF), Distance (PDIS) &	ICD, SED, SCD, Supplementary XML &	MMS, GOOSE,
R-GOOSE, R-SV\\
    \hline
  \end{tabular}
  \label{tabprot}
\end{footnotesize}
\end{table*}

\subsubsection{Configuring Cyber Network Emulator}\label{cybernetworkemulation}
In this section, the approach to derive the network topology to be emulated from an SCD file is elaborated. In order to generate cyber network emulation model, the topology of communication network has to be defined. This topology includes device addresses as well as connectivity among the devices. According to IEC 61850 standards, the information is defined in an SCD file. 

One of the parts of an SCD file is the `Communication' section, which includes details of subnetworks and access points, which correspond to devices in the process. The `SubNetwork' element contains `ConnectedAP'. As depicted in Fig.~\ref{commemulation}, it contains `Address' section that defines network address, net mask, and gateway information of each device (e.g., IED, PLC, and SCADA HMI). Another important section in the `ConnectedAP' is `PhysConn', which represents physical connectivity among devices. This section includes information about (physical) network ports on the device and the cable connected to each port. This cable name is unique within the SCD file. Therefore, the connectivity between devices and network switches and also between the network switches can be derived by utilizing this cable identification. For instance, in the example shown in Fig.~\ref{commemulation}, the same cable name is found in both IED and switch, and thus, a link between them is defined on the network emulator. The overall framework for emulating the communication section is similar to that of the power system emulation.

\subsubsection{Configuring Virtual Smart Grid Devices}\label{virtualied}
Another type of IEC 61850 SCL file, namely ICD file, can be utilized to configure and emulate virtual smart grid devices. The main section of the ICD file is the ‘IED’ section. This section contains the Logical Devices (LDs) and the Logical Nodes (LNs). LNs in IEC 61850 define the basic functions of an IED and they contain a group of Data Objects (DOs). IEC 61850 part 7-4 defines LN type or class~\cite{iecstd}.

\begin{figure}[b]
\vspace{-3mm}
\centerline
{\includegraphics[width=\linewidth]{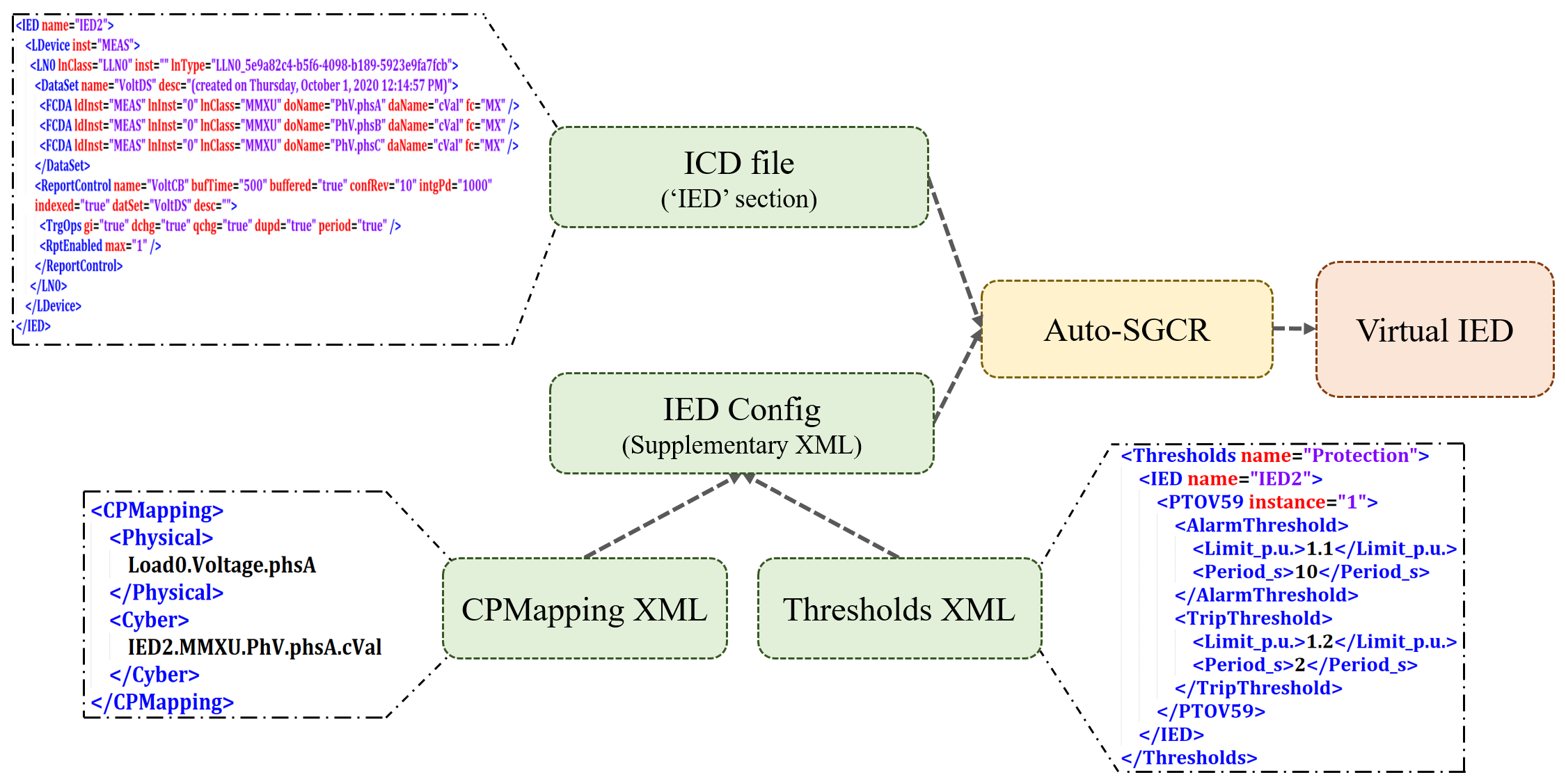}}
\caption{Virtual IED Implementation Framework.}
\label{icdsgmlvsd}
\end{figure}

In IEC 61850 ICD files, the protection LN class and definitions of dataset and attribute are important to configure virtual IEDs. Based on the number of substations, additional protection LNs have to be included in the virutal IEDs. The details of the protection logics implemented are tabulated in Table.~\ref{tabprot}. The popular protection functions include Over-current protection (PTOC), Over-voltage/Under-voltage protection (PTOV/PTUV), Under-voltage protection (PTUV), Inter-tripping (PTRC), Inter-locking (CILO), Differential protection (PDIF) and Distance Protection (PDIS). Presence of these LN elements indicates that the corresponding protection functions are implemented on an IED. An ICD file also enumerates DOs/attributes that are defined on each IED for monitoring and controlling power grid devices, including power grid measurements and device status (e.g., CB open/close). For instance, the measurements of three-phase voltages, three-phase currents, active power and frequency contribute towards the measurement LN class, represented as ‘MMXU’, and the attributes related to a CB are stored under `XCBR' LN class. A single IED may contain  instances of any LN type/class and this is represented by an instance number added as a suffix (e.g., MMXU1, MMXU2). Furthermore, ICD file defines datasets, each of which defines a set of data attributes exchanged or accessed together (e.g., in periodic reporting to PLCs/SCADA HMI and/or status update among IEDs). Additionally, ICD contains definitions of control blocks for IEC 61850 protocols. Such definitions are utilized by Auto-SGCR to configure communication functionality of each IED. Fig.~\ref{icdsgmlvsd} shows the overview of the framework for virtual IED configuration. As depicted, the inputs to the virtual IED configuration are the ICD files (mainly utilizing `IED' section) and Supplementary XML files. 

\begin{figure}[b]
\vspace{-4mm}
\centerline
{\includegraphics[width=1\linewidth, height=5.3cm]{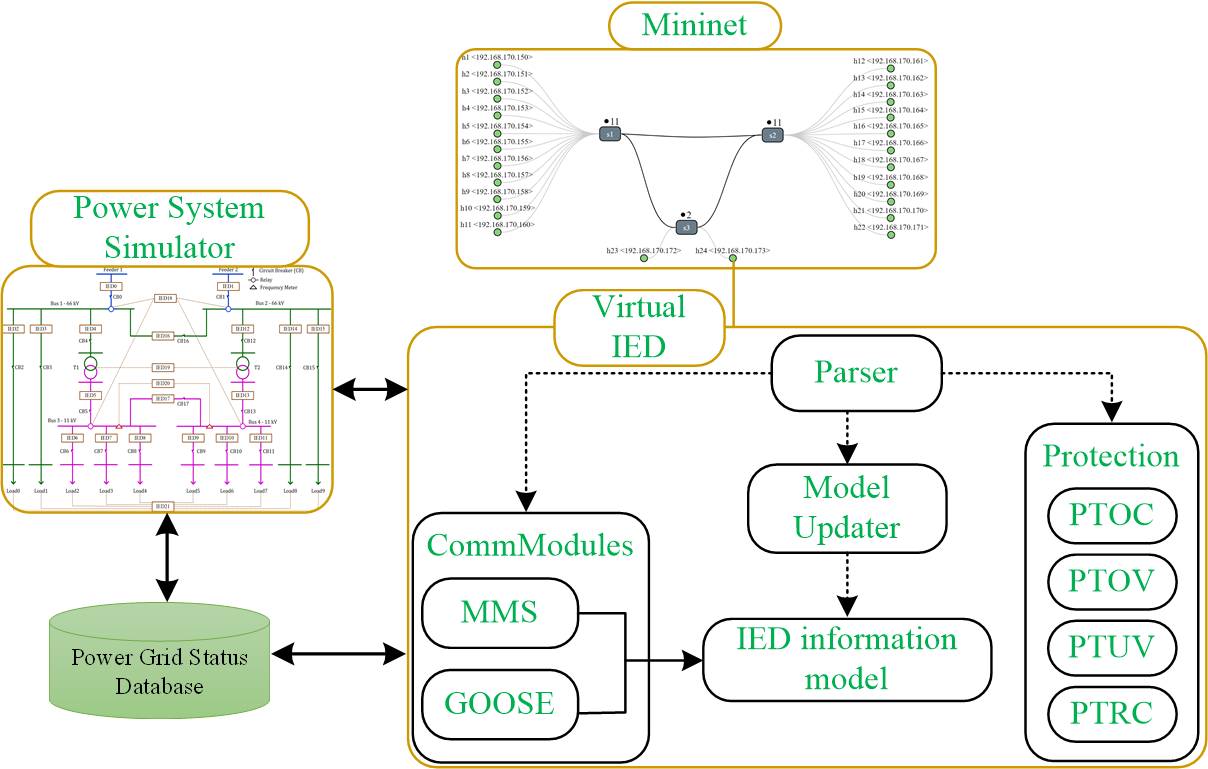}}
\caption{Cyber Range Generated by Auto-SGCR.}
\label{implem}
\end{figure}

\begin{figure}[b]
\vspace{-4mm}
\centerline
{\includegraphics[width=0.9\linewidth, height=5.3cm]{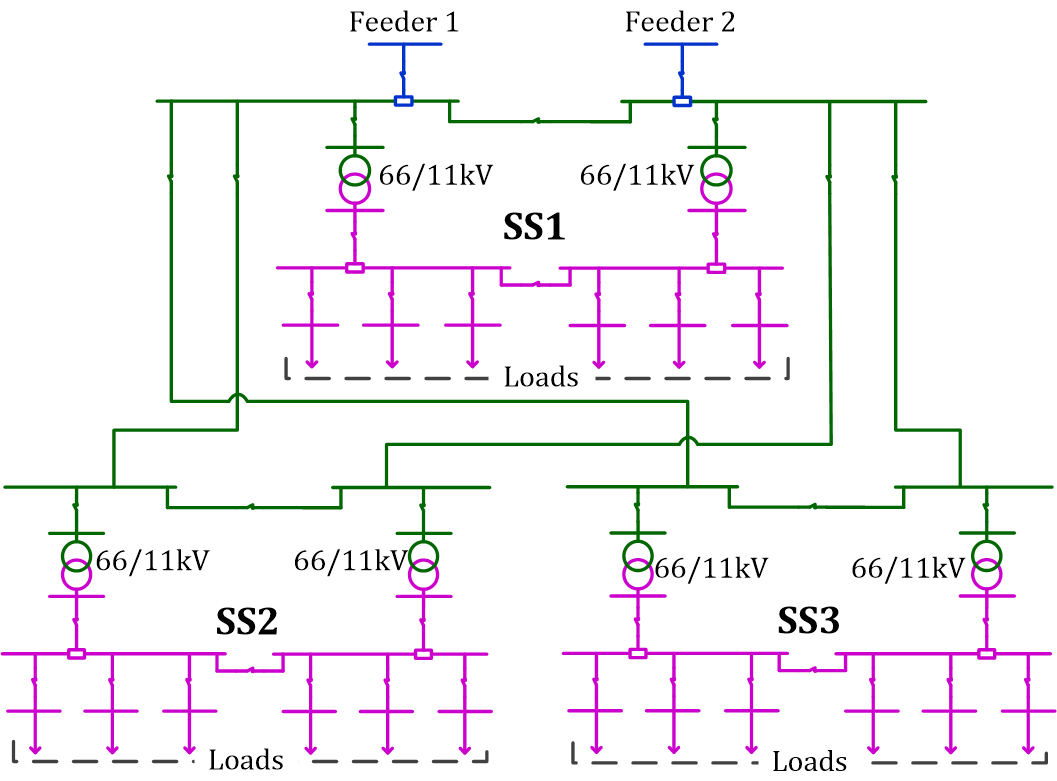}}
\caption{Single Line Diagram of 3-substation Model.}
\label{sld3}
\end{figure}

\begin{figure*}[hbt!]
  \centering
  \begin{subfigure}[b]{0.5\textwidth}
    \begin{center}
    {\includegraphics[width=1\linewidth, height=6cm]{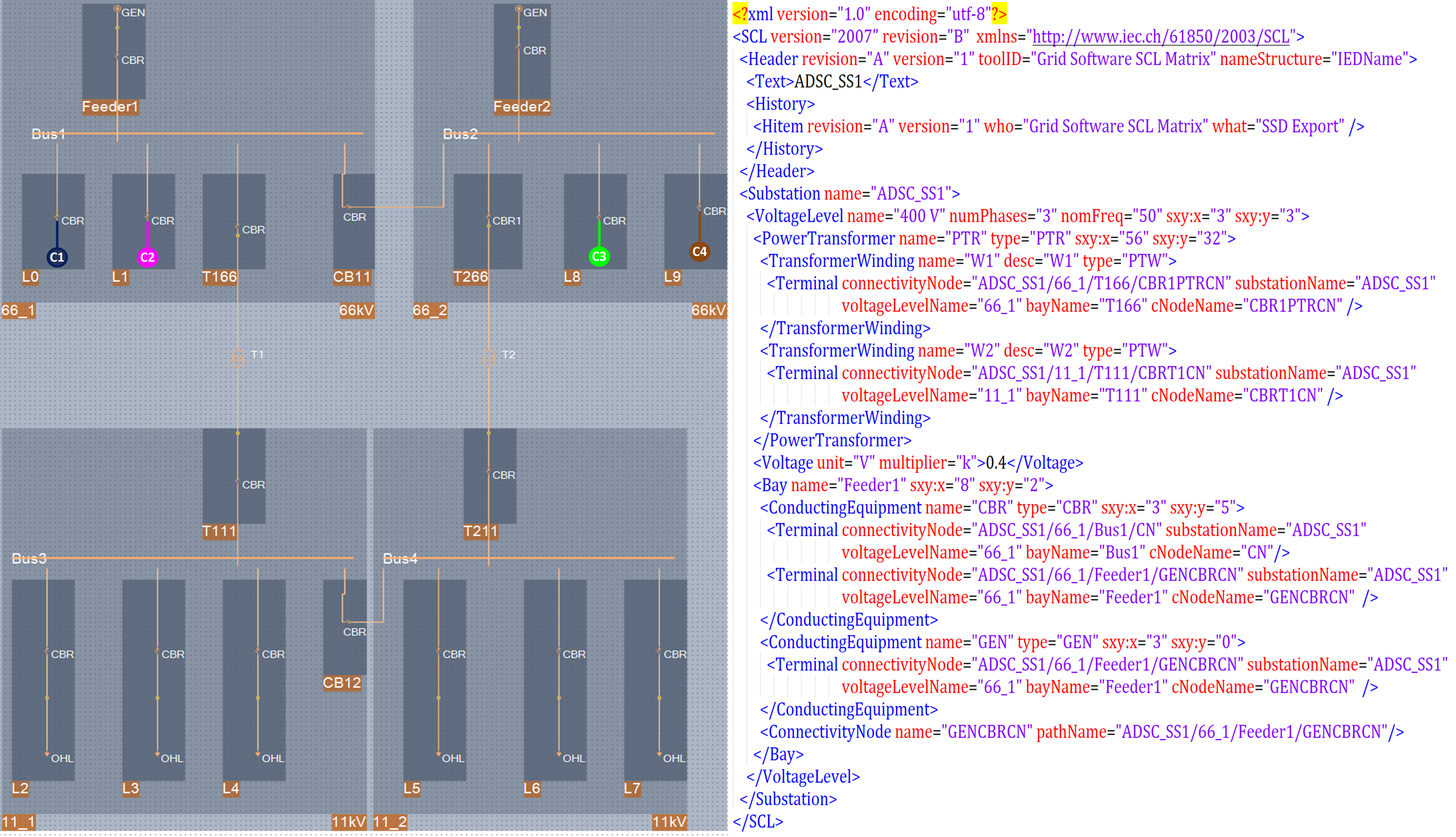}}
    \end{center}
    \caption{Input SSD (Physical Layout) of large scale substation.}
    \label{lssssdinput}
  \end{subfigure}
  \hspace{-1mm}
  \begin{subfigure}[b]{0.4\textwidth}
  \begin{center}
    {\includegraphics[width=\linewidth, height=6cm]{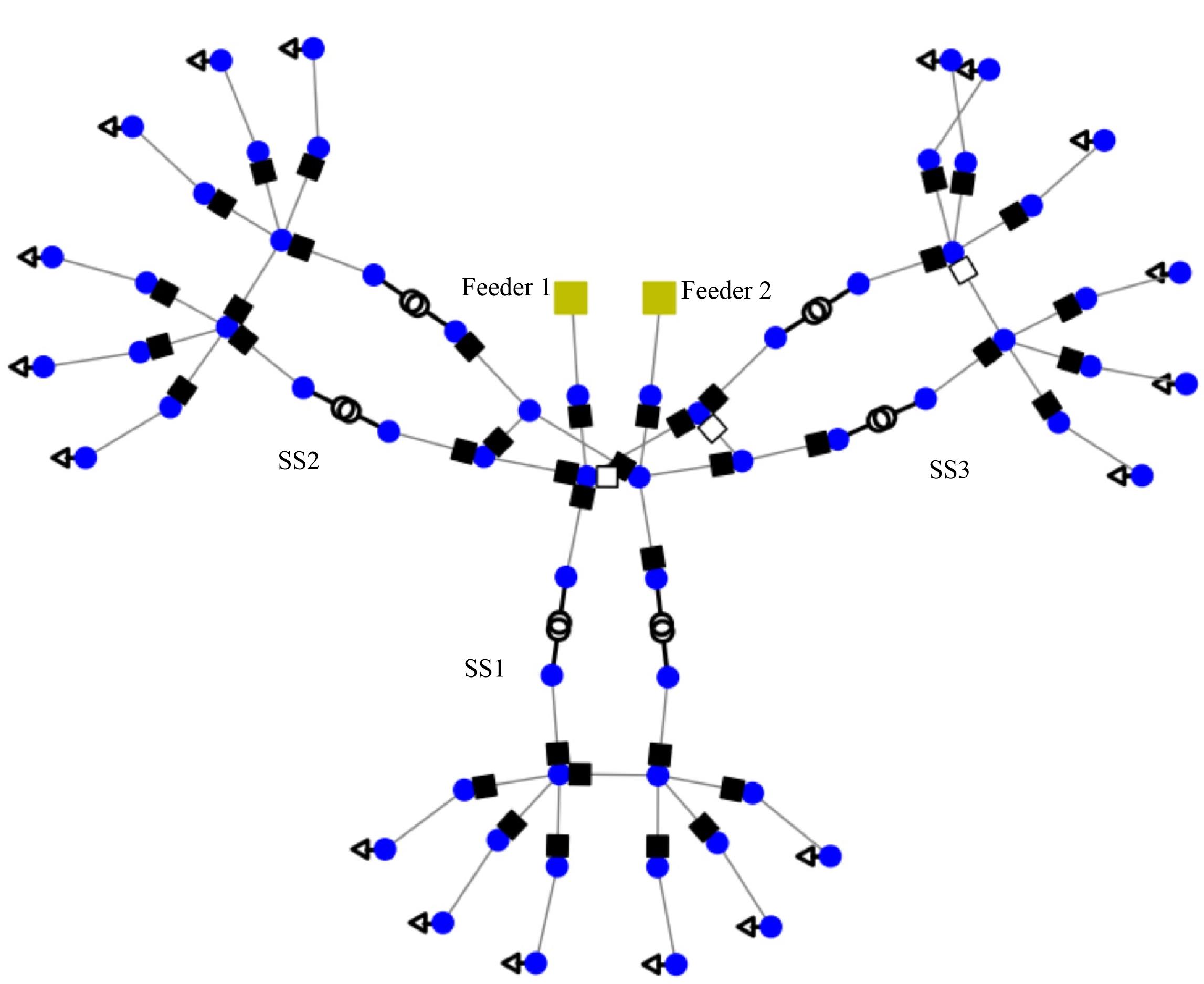}}
    \end{center}
    \caption{Output physical model generated by Auto-SGCR.}
    \label{lssssdoutput}
  \end{subfigure}
  \caption{Power System Topology Generation Using Auto-SGCR.}
  \vspace{-4.5mm}
\end{figure*}

Virtual IEDs are configured with extracting the relevant LN definitions, data attributes and dataset definitions, and communication/report control blocks in an ICD file. Based on these configurations, necessary protection functions and communication modules are loaded in the executable virtual IED. The CPMapping XML file is referred to when the virtual IED reads/writes data from the database to interact with the power system simulator. A snippet of the schema for mapping the physical measurements (\textit{Load0.Voltage.phsA}) with the IED attributes (\textit{IED2.MMXU.PhV.phsA.cVal}) is demonstrated in the CPMapping XML in Fig.~\ref{icdsgmlvsd}. 
 
The `Thresholds XML' file contains the threshold values for the virtual IED to execute the protection functions. In Fig.~\ref{icdsgmlvsd}, the threshold conditions for over-voltage protection (PTOV59) are exemplified. When a measurement of interest exceeds `Alarm threshold', the virtual IED sends alarm message; while the virtual IED immediately trips a CB if `Trip threshold' is violated. Similarly, the conditions for other protection functions are also defined.

The virtual PLC is implemented using an open-source software, named OpenPLC~\cite{openplc}. OpenPLC complies with IEC 61131-standard and is widely adopted to emulate the functionality of a PLC. The logic to be executed on the OpenPLC can be configured with PLCopen XML. However, as IEC 61850 standards are implemented in IEDs and the existing OpenPLC has the limitation of supporting the IEC 61850 standards, an enhanced version of OpenPLC named OpenPLC61850 was developed and published as an open-source project~\cite{softwarex}. This software supports Manufacturing Message Specification (MMS) based IED 61850 communication. This support is implemented using libiec61850~\cite{libiec}. Thus, a virtual PLC can be instantiated by deploying OpenPLC61850 on a virtual network node on the cyber network emulator. Network addresses and related information can be derived from an ICD file, just like virtual IEDs.

\subsubsection{Instantiating Smart Grid Cyber Range}\label{implementation}

The components of a smart grid cyber range generated by Auto-SGCR is illustrated in Fig.~\ref{implem}. Power system model that is configured as discussed in~\ref{powersystememulation} is run on Pandapower~\cite{panda} using Python~\cite{python}. The power flow information such as measurements and status are stored in MySQL~\cite{mysql} database. Subsequently, these data are read/updated by virtual IEDs, which are implemented with libiec61850, to process IEC 61850 information models and protocols. As discussed in subsection~\ref{cybernetworkemulation}, necessary modules (e.g., protection function and communication protocols) are loaded according to ICD and other configuration files. The cyber connectivity among devices (IEDs, PLCs, and switches) are configured as discussed in~\ref{cybernetworkemulation}) and translated into a network topology on Mininet emulator~\cite{mininet}.

The proof-of-concept implementation is done solely by using open-source software, which contributes accessibility to a broader user base as well as less complication when it is used for cybersecurity training and competition for public. Regardless, the cyber range and Auto-SGCR can be potentially extended to incorporate commercial simulators (e.g., Matlab Simulink) for better fidelity and realtimeness, which is on our road map.   

\section{Demonstration of Auto-SGCR}\label{demonstration}

In this section, the modeling and generation of cyber range for a large scale sub-transmission system is presented to demonstrate the usage and capability of the Auto-SGCR. 
The sub-transmission system consists of multi-substations model with three interconnected 66/11 kV smart substations connected to two incoming feeders. The single line diagram of the 3-substation model considered for this study is depicted in Fig.~\ref{sld3}. This evaluation is conducted to demonstrate the flexibility of Auto-SGCR in developing cyber range of real-world smart grid model. 

\subsection{SSD Translation}\label{demossd6611}

As aforementioned, the input to the Auto-SGCR are the SSD and SED files. Therefore, the corresponding SSD files for the multi-substation model is created using SCL matrix~\cite{sclmatrix}. A large scale multi-substation model is chosen to demonstrate the scalability/flexibility of the Auto-SGCR. The individual grid physical layout is defined in the respective SSD file. Fig.~\ref{lssssdinput} depicts the SLD of the first substation in the multi-substation model and a snippet of the SSD file. The details of the physical interconnection between the three substations are included in the SED file. Therefore, the SED file and SSDs of 3 substations are merged using the merger tool and the merged file is utilized as input to the Auto-SGCR tool. As these files, provide only the details regarding the physical layout, supplementary XML files that contain the specification of these physical components are also included as a part of input. 

The output of the large scale model generated using Auto-SGCR is portrayed in Fig.~\ref{lssssdoutput}. As represented in the figure, Auto-SGCR outputs the physical layout of the multi-substation model along with the configurations. For instance, the SCL files include the location of the CB in the substation. However, the status of the CB (if the CB is `open' or `close') is not defined. In Fig.~\ref{lssssdoutput}, the statuses of CBs are clearly indicated through the extra configuration XML files. Similarly, the specification of the feeders, transmission lines, transformer ratings and the load profiles are also fed through the XML files. 

\subsection{ICD Translation}\label{demoicd6611}

\begin{figure}[!tbp]
  \centering
  \begin{subfigure}[b]{0.45\textwidth}   
      \begin{center}
    {\includegraphics[width=\linewidth, height=6cm]{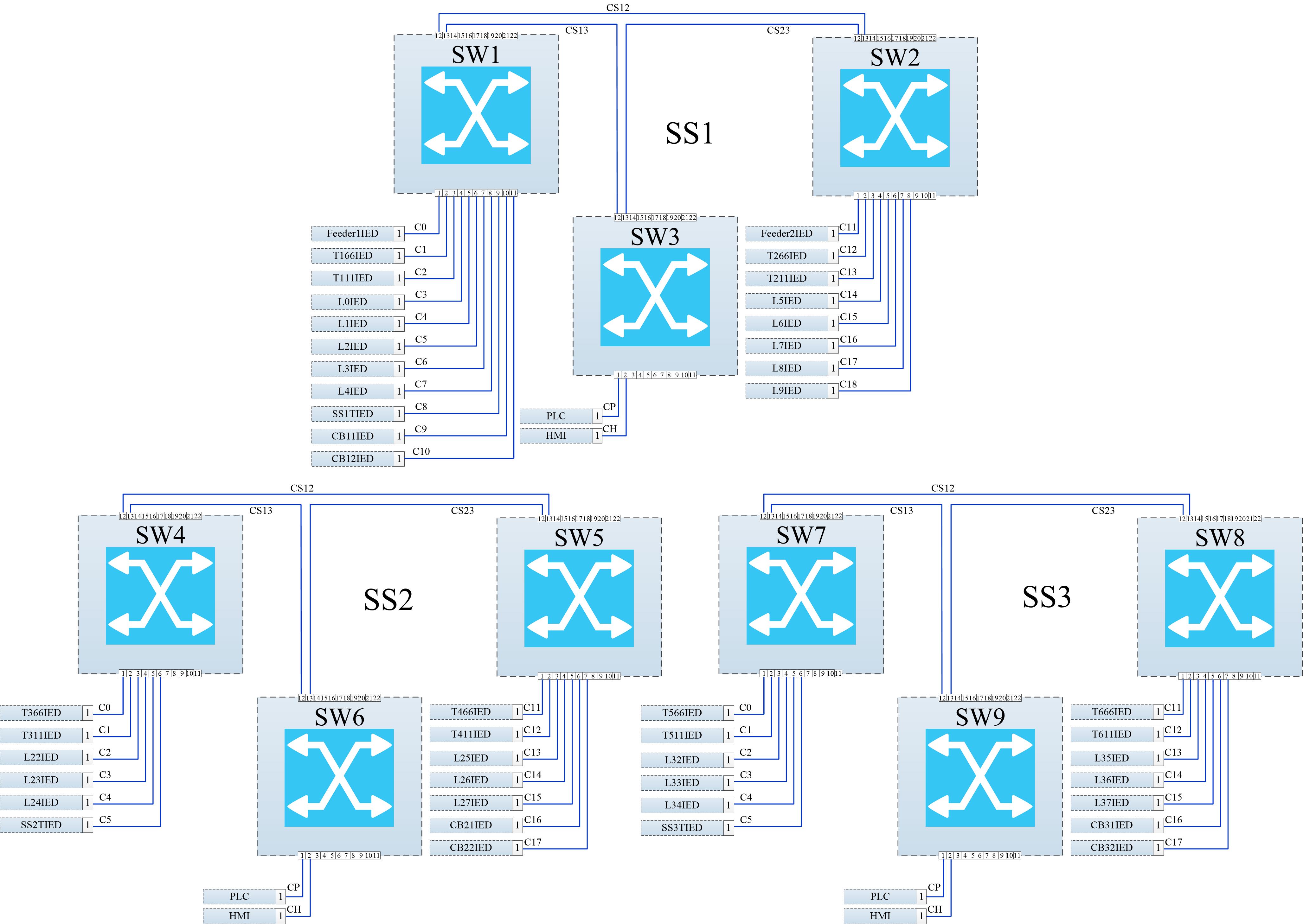}}
    \end{center}
    \caption{Input SCD (Communication Layout) of large scale model.}
    \label{lssscdinput}
  \end{subfigure}
  \hspace{-1mm}
  \begin{subfigure}[b]{0.45\textwidth}   
  \begin{center}      
{\includegraphics[width=\linewidth, height=6cm]{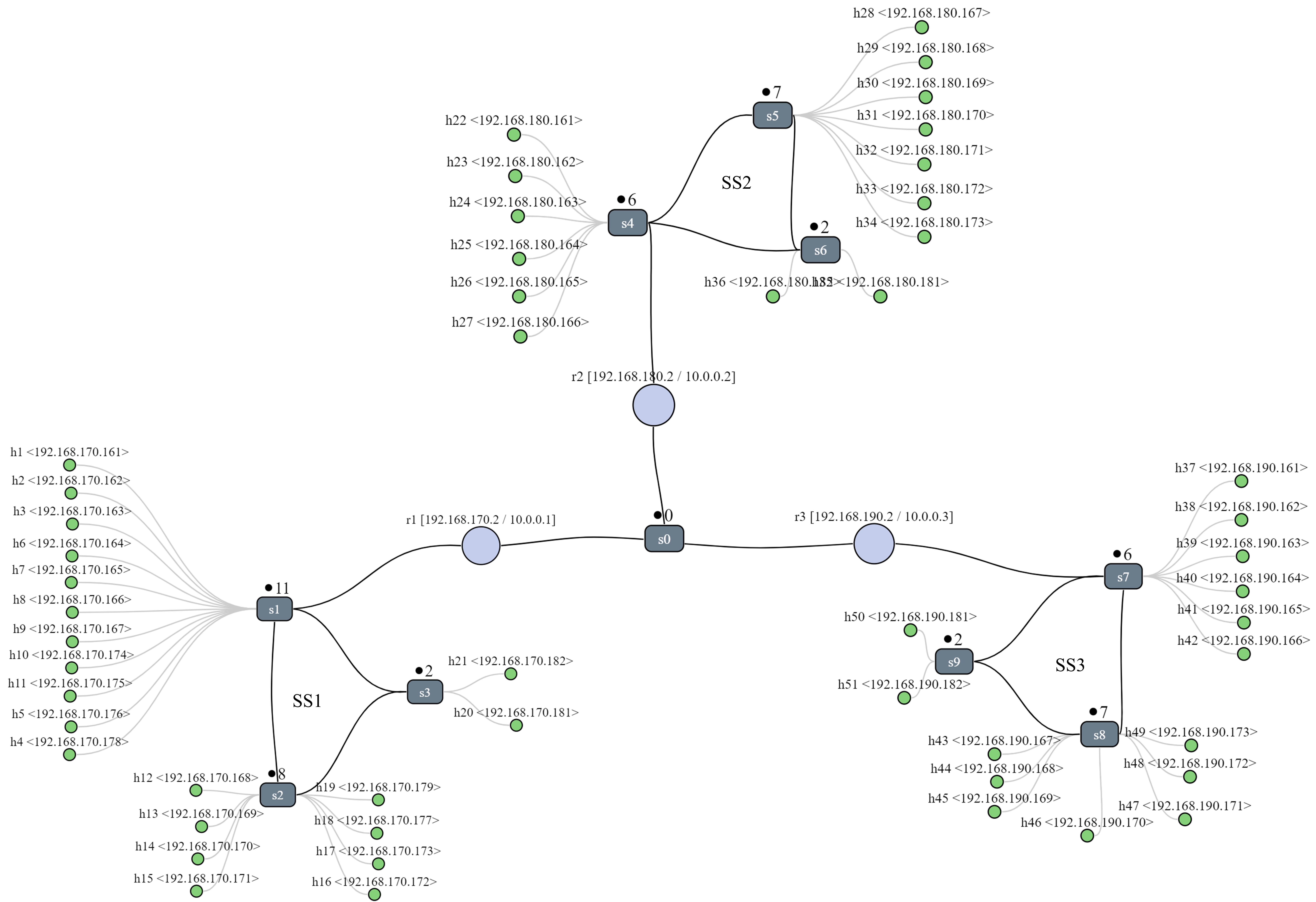}}
  \end{center}
    \caption{Output network model generated by Auto-SGCR.}
    \label{lssscdoutput}
  \end{subfigure}
    \caption{Cyber System Topology Generation Using Auto-SGCR.}
    \vspace{-3mm}
\end{figure}

As this model involves multiple substations, the number of IEDs involved is also higher. All the ICDs (for 45 IEDs) are created using SCL Matrix. The communication diagram is depicted in Fig.~\ref{lssscdinput}. The connection between different IEDs and switches in the same substation is defined in their respective SCD file. The SCD files from different substations are merged into a single SCD file, as explained in section~\ref{cnc}. This merged SCD file defines the physical connections among the switches in the same grid and between IEDs from multiple generation grids. This file is used as input to the Auto-SCGR and the output from the Auto-SGCR that illustrates all the IEDs, IEDs-switch, switch-switch connections is exemplified in Fig.~\ref{lssscdoutput}.

\subsection{Computational Performance}\label{performance}
One advantage of smart grid cyber range is scalability. Thus, this section presents the evaluation in terms of computational resources consumed for deploying the cyber range generated by Auto-SGCR. The evaluation is performed on a machine running Linux OS equipped with Intel i9 CPU and 16GB RAM. The resource consumption for small scale (9 IEDs) and large scale (45 IEDs, 104 IEDs) smart grid cyber range are evaluated. Both cyber range consist of multiple modules such as Pandapower, MySQL, virtual IED, Mininet, OpenPLC61850, and SCADA. Among them, the most resource demanding modules, namely the database and virtual IED are focused. Virtual IEDs periodically access database for reading and writing power grid status. 100ms interval and 1s interval are considered for the evaluation as these intervals roughly corresponds to the time taken by each iteration of power flow simulation and DB update on the same machine. The average memory and CPU consumption are summarized in Table~\ref{memory}. From the table, it can be noticed that the virtual IED's resource consumption increases linearly as expected. Based on the tabulated measurements, the DB polling interval (100 ms) for large scale model with 104 IEDs (virtual nodes) is close to the limit of the computer used for the experiment. However, the cyber range can span over multiple physical nodes. For instance, in the Mininet topology hosted by each physical node, the switch representing WAN can be mapped to the physical network interface card of the node. Subsequently, through the physical network connecting the physical nodes, Mininet virtual network on different physical nodes can communicate with each other with appropriate routing configurations. Thus, the scalability of the Auto-SGCR can be extended beyond the tabulated scale by using more machines. \\

\begin{table}[b]
\vspace{-3mm}
\caption{Computational resources for deploying Auto-SGCR 
}
\begin{footnotesize}
  \begin{tabular}{|p{.75cm}|p{0.7cm}|p{0.85cm}|p{0.4cm}|p{0.85cm}|p{0.5cm}|p{0.85cm}|p{0.5cm}|}
    \hline
    \multirow{2}{*}{\textbf{Interval}} &
    \multirow{2}{*}{\textbf{Module}} &
    \multicolumn{2}{l|}{\textbf{9 nodes}} &     \multicolumn{2}{l|}{\textbf{45 nodes}} &
    \multicolumn{2}{l|}{\textbf{104 nodes}}
    \\\cline{3-8}
    & & 
    \textbf{Memory} & \textbf{CPU} & 
    \textbf{Memory} & \textbf{CPU} & 
    \textbf{Memory} & \textbf{CPU}\\
    \hline
    \multirow{2}{*}{100ms} & DB & 446MB &5.4\% & 514MB & 19.8\% & 594MB & 28.6\% \\
    \cline{2-8}
    & IED & 120MB & 4.0\% & 521MB & 17.6\% & 1GB & 36.8\% \\
    \hline
    \multirow{2}{*}{1s} & DB & 447MB &0.5\% & 511MB & 1.7\% & 585MB & 4.8\% \\
    \cline{2-8}
    & IED & 96MB & 2.5\% & 463MB & 7.4\% & 1GB & 18.7\% \\
    \hline
  \end{tabular}
  \label{memory}
\end{footnotesize}
\end{table}

\vspace{-3mm}
\section{Cyber-attack Case Study }\label{casestudy}
In this section, a realistic cyber attack experiment that can be conducted using the generated smart grid cyber range is discussed. Inherent vulnerabilities in smart grid communication is due to the lack of reliable sender and message authentication~\cite{fpro}. While IEC 62351 has been established to define cyber security for IEC 61850 protocols~\cite{iec62351}, it has not been widely deployed yet~\cite{babu2017security}. This is ideal for an adversary to carry out injection of malicious control command to IEDs~\cite{acmd-tsg, roomiaccess}. Thus, the FCI attack against inter-substation communication using IEC 61850 R-GOOSE protocol is discussed. As R-GOOSE is based on UDP multicast, spoofing of a sender identity is feasible. 

{An attacker can launch false command to the network either by impersonating a legitimate IED and providing malicious commands to the other IED or by compromising any IED in the network and modify the values reported by the IED. In this paper, a FCI attack is demonstrated on the sub-transmission model generated by Auto-SGCR in section~\ref{demonstration}. The substation 1 and substation 2 in the sub-transmission system shown in Fig.~\ref{sld3} are connected by a feeder. The interconnecting feeder has two IEDs: `S1\_IED22' in substation 1 and `S2\_IED0' in substation 2 monitoring the CBs at both ends. The CBs of the interconnecting feeders are interlocked. The pseudocode in Algorithm 1 illustrates the interlocking mechanism implemented between S1\_IED22 and S2\_IED0. 

\footnotesize{\textbf{Algorithm 1:} Interlocking Mechanism\\}
\tab \footnotesize{if (S1\_IED22\$XCBR.Pos):}\\
\tab\tab \footnotesize{ S2\_IED0\$XCBR.Pos = 1  // Close}\\
\tab else\\
\tab\tab \footnotesize{ S2\_IED0\$XCBR.Pos = 0  // Open\\}}
\vspace{-2mm}

\begin{figure}[t]
\centerline
{\includegraphics[width=0.9\linewidth, height=4.5cm]{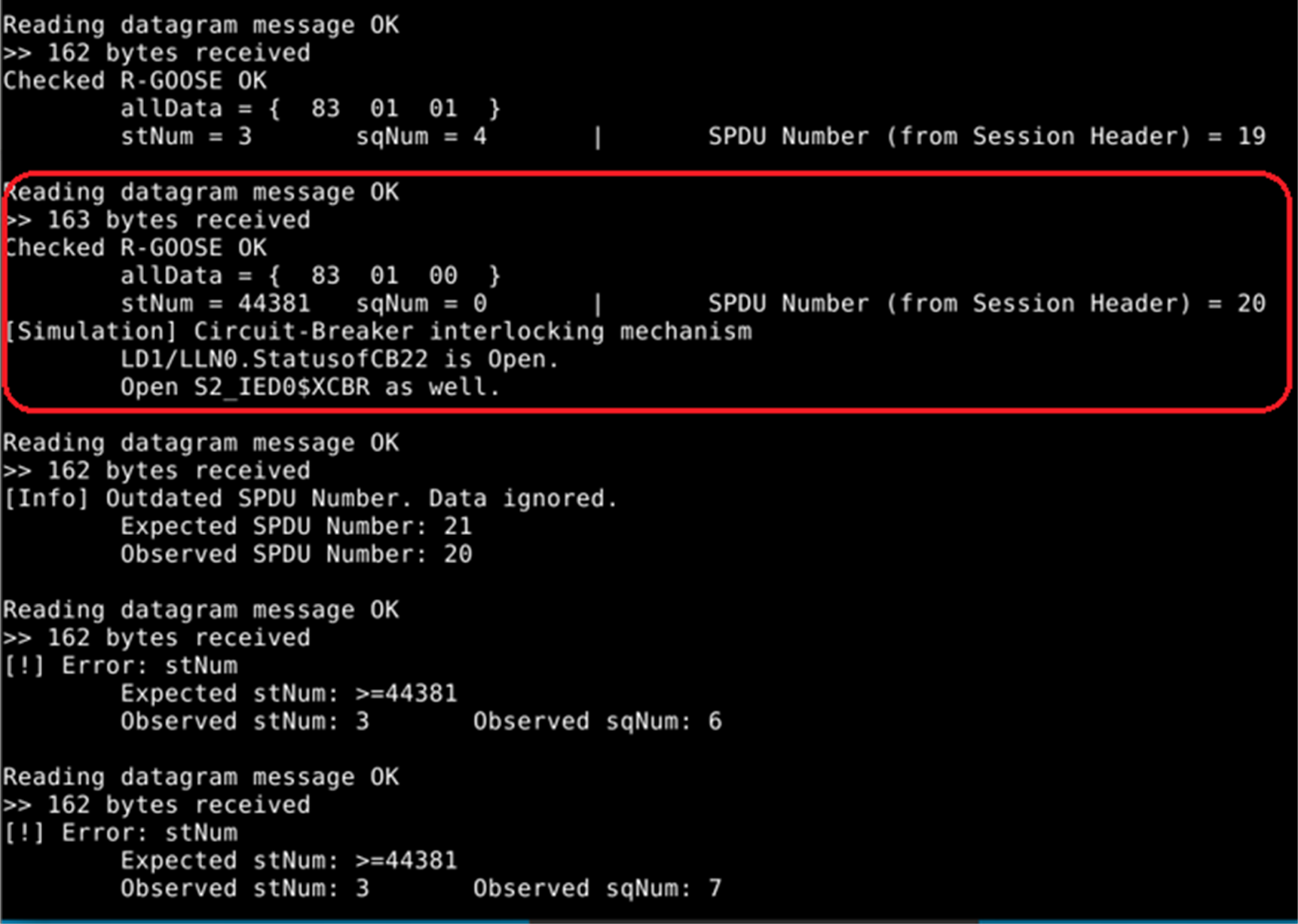}}
\caption{Screenshot of S2\_IED0 During Spoofed `stNum' FCI Attack.}
\label{fig:fci}
\vspace{-3mm}
\end{figure}

The status information of S1\_IED22 is communicated to S2\_IED0 through R-GOOSE messages. The R-GOOSE messages contains a counter `stNum' which is incremented whenever there is change in R-GOOSE data field. The subscriber IED, which subscribes to the R-GOOSE message, accepts the only if the current `stNum' is equal to or greater than the previous messages `stNum'. This is done to avoid acceptance of any delayed re-transmitted message and mitigate replay attacks~\cite{stnum}. The attacker eavesdrops on the network, and when the status of S1\_IED22 is `close', the attacker injects a R-GOOSE packet with a large `stNum' value. This causes that the legitimate IED's message would not be accepted by the subscriber. Fig.~\ref{fig:fci} show that subsequent real messages from the original R-GOOSE publisher were ignored by the subscriber, due to a smaller-than-expected `stNum'. Subsequently, the attacker could inject a fake R-GOOSE message indicating that the circuit breaker is `open'. When the attacker’s packet arrived at the S2\_IED0, the spoofed status is accepted and passed to the Protection Logic. This triggered S2\_IED0\$XCBR to open, as highlighted in the boxed-up portion of the console output in Fig.~\ref{fig:fci}. 

\section{Conclusions}\label{conclusion}

In order to design and operate dependable smart grid systems, conducting extensive cyber-attack experiments to evaluate the robustness of the system is essential. Smart grid cyber range is a viable solution. However, development of such systems requires intensive domain knowledge in both physical and network aspects. Besides, the existing smart grid cyber range efforts are either proprietary that limits accessibility, or designed as an one-off solution that lacks customizability. Thus, a cyber range that meets individual user's preference or requirement still remains a challenge.

In this paper, a framework to automate the generation of smart grid cyber range, called Auto-SGCR is developed. The framework consists of human-/machine-friendly IEC 61850 standard based modelling language. The application of the Auto-SGCR is demonstrated by instantiating cyber range for multi-substation smart grid model. Case studies of representative cyber-attack strategies against smart grid systems are also conducted. The schema and the Auto-SGCR toolchain to process the model is open-sourced. The developed framework allows power grid operators to utilize their existing IEC 61850 based configurations and also other users to generate and customize their cyber range without requiring intensive efforts. Moreover, these models can be easily shared in the user community to reproduce the cyber range.  
\balance
\bibliographystyle{IEEEtran}
\bibliography{references}

\begin{thebibliography}{10}
\providecommand{\url}[1]{#1}
\csname url@samestyle\endcsname
\providecommand{\newblock}{\relax}
\providecommand{\bibinfo}[2]{#2}
\providecommand{\BIBentrySTDinterwordspacing}{\spaceskip=0pt\relax}
\providecommand{\BIBentryALTinterwordstretchfactor}{4}
\providecommand{\BIBentryALTinterwordspacing}{\spaceskip=\fontdimen2\font plus
\BIBentryALTinterwordstretchfactor\fontdimen3\font minus \fontdimen4\font\relax}
\providecommand{\BIBforeignlanguage}[2]{{%
\expandafter\ifx\csname l@#1\endcsname\relax
\typeout{** WARNING: IEEEtran.bst: No hyphenation pattern has been}%
\typeout{** loaded for the language `#1'. Using the pattern for}%
\typeout{** the default language instead.}%
\else
\language=\csname l@#1\endcsname
\fi
#2}}
\providecommand{\BIBdecl}{\relax}
\BIBdecl

\bibitem{ANCILLOTTI20131665}
E.~Ancillotti, R.~Bruno, and M.~Conti, ``{The role of communication systems in smart grids: Architectures, technical solutions and research challenges},'' \emph{Computer Communications}, vol.~36, pp. 1665--1697, 2013.

\bibitem{aftab20}
M.~A. Aftab, S.~S. Hussain, I.~Ali, and T.~S. Ustun, ``{IEC 61850 based substation automation system: A survey},'' \emph{International Journal of Electrical Power \& Energy Systems}, vol. 120, p. 106008, 2020.

\bibitem{SUN201845}
C.-C. Sun, A.~Hahn, and C.-C. Liu, ``{Cyber security of a power grid: State-of-the-art},'' \emph{International Journal of Electrical Power \& Energy Systems}, vol.~99, pp. 45--56, 2018.

\bibitem{8731}
V.~Pillitteri and T.~Brewer, ``\BIBforeignlanguage{en}{{Guidelines for Smart Grid Cybersecurity}},'' 2014.

\bibitem{idaho}
``{Aurora Vulnerability: Origin, Explanation and Solutions},'' Available at \url{https://tinyurl.com/fneh32ap}.

\bibitem{davis2013survey}
J.~Davis and S.~Magrath, ``A survey of cyber ranges and testbeds,'' 2013.

\bibitem{yamin2020cyber}
M.~M. Yamin, B.~Katt, and V.~Gkioulos, ``Cyber ranges and security testbeds: Scenarios, functions, tools and architecture,'' \emph{Computers \& Security}, vol.~88, p. 101636, 2020.

\bibitem{dsn2023}
D.~Mashima, S.~M.~M. Roomi, B.~Ng, Z.~Kalberczyk, S.~Hussain, and E.-C. Chang, ``Towards automated generation of smart grid cyber range for cybersecurity experiments and training,'' in \emph{Proceedings of DSN 2023 Industry Track}, 2023.

\bibitem{hardwarecps}
A.~Sahu, P.~Wlazlo, Z.~Mao, H.~Huang, A.~Goulart, K.~Davis, and S.~Zonouz, ``Design and evaluation of a cyber-physical testbed for improving attack resilience of power systems,'' \emph{IET Cyber-Physical Systems: Theory \& Applications}, vol.~6, no.~4, pp. 208--227, 2021.

\bibitem{cosim3}
C.~M. Davis, J.~E. Tate, H.~Okhravi, C.~Grier, T.~J. Overbye, and D.~Nicol, ``Scada cyber security testbed development,'' in \emph{2006 38th North American Power Symposium}, 2006, pp. 483--488.

\bibitem{2019817}
E.~Hammad, M.~Ezeme, and A.~Farraj, ``Implementation and development of an offline co-simulation testbed for studies of power systems cyber security and control verification,'' \emph{International Journal of Electrical Power \& Energy Systems}, vol. 104, pp. 817--826, 2019.

\bibitem{cosim5}
Z.~Liu, Q.~Wang, Y.~Tang, and M.~Ni, ``The real-time co-simulation platform with hardware-in-loop for cyber-attack in smart grid,'' in \emph{IEEE Innovative Smart Grid Technologies (ISGT Asia)}, 2018, pp. 845--849.

\bibitem{cosim6}
K.~Pan, A.~Teixeira, C.~D. López, and P.~Palensky, ``Co-simulation for cyber security analysis: Data attacks against energy management system,'' in \emph{2017 IEEE International Conference on Smart Grid Communications (SmartGridComm)}, 2017, pp. 253--258.

\bibitem{cosim7}
V.~Venkataramanan, A.~Srivastava, and A.~Hahn, ``Real-time co-simulation testbed for microgrid cyber-physical analysis,'' in \emph{2016 Workshop on Modeling and Simulation of Cyber-Physical Energy Systems (MSCPES)}, 2016, pp. 1--6.

\bibitem{cosim8}
Y.~Cao, X.~Shi, Y.~Li, Y.~Tan, M.~Shahidehpour, and S.~Shi, ``A simplified co-simulation model for investigating impacts of cyber-contingency on power system operations,'' \emph{IEEE Transactions on Smart Grid}, vol.~9, no.~5, pp. 4893--4905, 2018.

\bibitem{cosim9}
N.~Duan, N.~Yee, B.~Salazar, J.-Y. Joo, E.~Stewart, and E.~Cortez, ``Cybersecurity analysis of distribution grid operation with distributed energy resources via co-simulation,'' in \emph{2020 IEEE Power \& Energy Society General Meeting (PESGM)}, 2020, pp. 1--5.

\bibitem{cosim10}
N.~K. Kandasamy, S.~Venugopalan, T.~K. Wong, and N.~J. Leu, ``An electric power digital twin for cyber security testing, research and education,'' \emph{Computers and Electrical Engineering}, vol. 101, p. 108061, 2022.

\bibitem{cpssurvey}
A.~A. Smadi, B.~T. Ajao, B.~K. Johnson, H.~Lei, Y.~Chakhchoukh, and Q.~Abu Al-Haija, ``A comprehensive survey on cyber-physical smart grid testbed architectures: Requirements and challenges,'' \emph{Electronics}, vol.~10, no.~9, 2021.

\bibitem{DERtestbed}
H.~Albunashee, C.~Farnell, A.~Suchanek, K.~Haulmark, R.~McCann, J.~Di, and A.~Mantooth, ``A testbed for detecting false data injection attacks in systems with distributed energy resources,'' \emph{IEEE Journal of Emerging and Selected Topics in Power Electronics}, pp. 1--1, 2019.

\bibitem{scadatestbed}
I.~N. Fovino, M.~Masera, L.~Guidi, and G.~Carpi, ``An experimental platform for assessing scada vulnerabilities and countermeasures in power plants,'' in \emph{3rd International Conference on Human System Interaction}, 2010, pp. 679--686.

\bibitem{scadaidaho}
``{National SCADA Test Bed Substation Automation Evaluation Report},'' Available at \url{https:https://inldigitallibrary.inl.gov/sites/sti/sti/4374057.pdf}, 2009.

\bibitem{epic}
iTrust, Available at \url{https://itrust.sutd.edu.sg/testbeds/electric-power-intelligent-control-epic/}.

\bibitem{lowcost}
M.~Annor-Asante and B.~Pranggono, ``Development of smart grid testbed with low-cost hardware and software for cybersecurity research and education,'' \emph{Wireless Pers Commun}, vol. 101, p. 1357–1377, 2018.

\bibitem{2017124}
S.~Poudel, Z.~Ni, and N.~Malla, ``Real-time cyber physical system testbed for power system security and control,'' \emph{International Journal of Electrical Power \& Energy Systems}, vol.~90, pp. 124--133, 2017.

\bibitem{5759169}
M.~Mallouhi, Y.~Al-Nashif, D.~Cox, T.~Chadaga, and S.~Hariri, ``A testbed for analyzing security of scada control systems (tasscs),'' in \emph{ISGT}, 2011.

\bibitem{pw}
``{PowerWorld},'' Available at \url{https://www.powerworld.com/}.

\bibitem{fdimss}
M.~M. Roomi, P.~P. Biswas, D.~Mashima, Y.~Fan, and E.-C. Chang, ``{False Data Injection Cyber Range of Modernized Substation System},'' in \emph{2020 IEEE International Conference on Communications, Control, and Computing Technologies for Smart Grids}.\hskip 1em plus 0.5em minus 0.4em\relax IEEE, 2020, pp. 1--7.

\bibitem{7116592}
R.~Liu, C.~Vellaithurai, S.~S. Biswas, T.~T. Gamage, and A.~K. Srivastava, ``Analyzing the cyber-physical impact of cyber events on the power grid,'' \emph{IEEE Transactions on Smart Grid}, vol.~6, no.~5, pp. 2444--2453, 2015.

\bibitem{costCPS}
G.~Elbez, H.~B. Keller, and V.~Hagenmeyer, ``{A Cost-efficient Software Testbed for Cyber-Physical Security in IEC 61850-based Substations},'' in \emph{IEEE International Conference on Communications, Control, and Computing Technologies for Smart Grids}, 2018.

\bibitem{SoftGrid}
P.~Gunathilaka, D.~Mashima, and B.~Chen, ``Softgrid: A software-based smart grid testbed for evaluating substation cybersecurity solutions,'' in \emph{Proceedings of the 2nd ACM Workshop on Cyber-Physical Systems Security and Privacy}, New York, NY, USA, 2016, p. 113–124.

\bibitem{roomi2023analysis}
M.~M. Roomi, S.~S. Hussain, D.~Mashima, E.-C. Chang, and T.~S. Ustun, ``Analysis of false data injection attacks against automated control for parallel generators in iec 61850-based smart grid systems,'' \emph{IEEE Systems Journal}, vol.~17, no.~3, pp. 4603--4614, 2023.

\bibitem{61850-6}
``{Communication networks and systems for power utility automation - Part 6: Configuration description language for communication in power utility automation systems related to IEDs},'' \emph{IEC 61850-6:2009+AMD1:2018 CSV consolidated version}, 2018.

\bibitem{plcopen}
``Technical paper plcopen technical committee 6 - xml formats for iec 61131-3,'' Available at \url{https:https://www.plcopen.org/system/files/downloads/tc6\_xml\_v201\_technical\_doc.pdf}, year = {2009}.

\bibitem{iecstd}
``{IEC 61850-7-4.; Communication networks and systems for power utility automation – Basic communication structure – Compatible LN classes and DO classes, IEC International Standard, Ed. 2.0},'' 2010.

\bibitem{openplc}
T.~Alves and T.~Morris, ``{OpenPLC: An IEC 61,131--3 compliant open source industrial controller for cyber security research},'' \emph{{Computers \& Security}}, vol.~78, pp. 364--379, 2018.

\bibitem{softwarex}
{M. M. Roomi, W. S. Ong, D. Mashima, and S. S. M. Hussain}, ``{OpenPLC61850: An IEC 61850 MMS compatible open source PLC for smart grid research},'' \emph{SoftwareX}, vol.~17, p. 100917, 2022.

\bibitem{libiec}
``{libIEC61850},'' Available at \url{https://libiec61850.com/libiec61850/}.

\bibitem{panda}
``{pandapower},'' Available at \url{http://www.pandapower.org/}.

\bibitem{python}
``Python,'' Available at \url{https://www.python.org/}.

\bibitem{mysql}
``{MySQL},'' Available at \url{https://www.mysql.com/}.

\bibitem{mininet}
``{Mininet},'' Available at \url{http://mininet.org/}.

\bibitem{sclmatrix}
``{GridSoftware},'' Available at \url{http://www.gridsoftware.com/products/sclmatrix.html}.

\bibitem{fpro}
E.~Esiner, D.~Mashima, B.~Chen, Z.~Kalbarczyk, and D.~Nicol, ``F-pro: a fast and flexible provenance-aware message authentication scheme for smart grid,'' in \emph{2019 IEEE International Conference on Communications, Control, and Computing Technologies for Smart Grids}, 2019, pp. 1--7.

\bibitem{iec62351}
S.~M.~S. Hussain, T.~S. Ustun, and A.~Kalam, ``A review of iec 62351 security mechanisms for iec 61850 message exchanges,'' \emph{IEEE Transactions on Industrial Informatics}, vol.~16, no.~9, pp. 5643--5654, 2020.

\bibitem{babu2017security}
B.~Babu, T.~Ijyas, P.~Muneer, and J.~Varghese, ``Security issues in scada based industrial control systems,'' in \emph{2017 2nd International Conference on Anti-Cyber Crimes (ICACC)}.\hskip 1em plus 0.5em minus 0.4em\relax IEEE, 2017, pp. 47--51.

\bibitem{acmd-tsg}
D.~Mashima, P.~Gunathilaka, and B.~Chen, ``Artificial command delaying for secure substation remote control: Design and implementation,'' \emph{IEEE Transactions on Smart Grid}, vol.~10, no.~1, pp. 471--482, 2019.

\bibitem{roomiaccess}
M.~M. Roomi, W.~S. Ong, S.~M.~S. Hussain, and D.~Mashima, ``{IEC 61850 Compatible OpenPLC for Cyber Attack Case Studies on Smart Substation Systems},'' \emph{IEEE Access}, vol.~10, pp. 9164--9173, 2022.

\bibitem{stnum}
N.~Kush, E.~Ahmed, M.~Branagan, and E.~Foo, ``Poisoned goose: Exploiting the goose protocol,'' in \emph{Proceedings of the Twelfth Australasian Information Security Conference - Volume 149}, ser. AISC '14.\hskip 1em plus 0.5em minus 0.4em\relax AUS: Australian Computer Society, Inc., 2014, p. 17–22.

\end{thebibliography}

\end{document}